\documentclass[opre,nonblindrev]{informs3_hide}
\OneAndAHalfSpacedXI 

\usepackage[english]{babel}
\usepackage[autostyle, english = american]{csquotes}
\MakeOuterQuote{"}
\usepackage[colorlinks,linkcolor=blue, citecolor=blue]{hyperref}
\usepackage[normalem]{ulem}

\usepackage{cleveref}
\crefname{subsection}{section}{subsections}
\usepackage{eqnarray}
\usepackage{algorithm,algpseudocode}
\usepackage{amsmath, bbm, enumitem, xparse, bm, lipsum}
\usepackage{amssymb}

\newcommand{\bI}{\mathbbm{1}}

\newcommand{\ALG}{\mathsf{ALG}}
\newcommand{\OPT}{\mathsf{OPT}}
\newcommand{\INDEP}{\mathsf{INDEP}}
\newcommand{\CORREL}{\mathsf{CORREL}}

\newcommand{\bs}{\bm{s}}

\newcommand{\bc}{\bm{c}}

\newcommand{\ba}{\bm{a}}

\usepackage[dvipsnames]{xcolor}

\usepackage{xparse}

\NewDocumentEnvironment{myproof}{o}
{\IfNoValueTF{#1}{\paragraph{{Proof.} }} {\paragraph{{#1.} }} }
{\hfill$\Halmos$}

\usepackage{natbib,bbm,multirow,multicol}
 \bibpunct[, ]{(}{)}{,}{a}{}{,}%
 \def\bibfont{\small}%
\TheoremsNumberedThrough     
\ECRepeatTheorems

\EquationsNumberedThrough    

\begin{document}




\TITLE{
\Large Constant Approximation for Network Revenue Management with Markovian-Correlated Customer Arrivals
}

\ARTICLEAUTHORS{%
\AUTHOR{$\text{Jiashuo Jiang}$}

\AFF{\  \\
Department of Industrial Engineering \& Decision Analytics, Hong Kong University of Science and Technology\\
}
}

\ABSTRACT{The Network Revenue Management (NRM) problem is a well-known challenge in dynamic decision-making under uncertainty. In this problem, fixed resources must be allocated to serve customers over a finite horizon, while customers arrive according to a stochastic process. The typical NRM model assumes that customer arrivals are independent over time. However, in this paper, we explore a more general setting where customer arrivals over different periods can be correlated. We propose a model that assumes the existence of a system state, which determines customer arrivals for the current period. This system state evolves over time according to a time-inhomogeneous Markov chain. We show our model can be used to represent correlation in various settings.

To solve the NRM problem under our correlated model, we derive a new linear programming (LP) approximation of the optimal policy. Our approximation provides an upper bound on the total expected value collected by the optimal policy. We use our LP to develop a new bid price policy, which computes bid prices for each system state and time period in a backward induction manner. The decision is then made by comparing the reward of the customer against the associated bid prices. Our policy guarantees to collect at least $1/(1+L)$ fraction of the total reward collected by the optimal policy, where $L$ denotes the maximum number of resources required by a customer.

In summary, our work presents a Markovian model for correlated customer arrivals in the NRM problem and provides a new LP approximation for solving the problem under this model. We derive a new
bid price policy and provides a theoretical guarantee of the performance of the policy.

}



\maketitle

\section{Introduction}

Network revenue management (NRM) is a classical problem in stochastic dynamic decision-making with resource constraints. The NRM problem involves allocating $m$ resources, each with an individual initial capacity, over a finite time horizon of length $T$. At each time step $t=1,\ldots,T$, a customer $t$ arrives, belonging to a type $j_t\in[n]$, and demands a vector $\ba_t\in\{0,1\}^m$ of resources, with a corresponding reward $r_t$. Both $\ba_t$ and $r_t$ are dependent on the customer type. After observing $\ba_t$ and $r_t$, an irrevocable decision must be made about whether to serve customer $t$. If served, $\ba_t$ units are consumed from the resources, and a $r_t$ reward is collected. Customer $t$ can only be served if there are enough remaining resources, i.e., each remaining resource is no smaller than the corresponding component of the demanding vector $\ba_t$. Note that even if there are enough remaining resources, a customer can be intentionally rejected to save resources for serving customers with higher values that may arrive in the future. The goal of the decision-maker is to maximize the total reward collected from serving customers.

One crucial aspect of the NRM model is how the size and reward $(\ba_t, r_t)$ are generated, i.e., how the customer type is determined for each customer $t$. The NRM problem is typically studied under the stochastic setting where $(\ba_t, r_t)$ are drawn from a distribution. The broad literature on the NRM problem has focused on the \textit{independent} customer arrival models (e.g. \cite{gallego1994optimal}), where $(\ba_t, r_t)$ are drawn from a distribution (which can be time-inhomogeneous) independently for each period $t$. However, as noted in recent studies \citep{bai2022fluid, aouad2022nonparametric}, a major shortcoming of the independent customer arrival model is that it cannot handle the coexistence of large demand volume and large demand variability. Indeed, for the independent customer arrival model, over the entire horizon, the variance of the type $j$ customer arrivals cannot exceed its expectation, which can be violated in practice. Additionally, demand can be non-stationary and can evolve over time in many business settings due to factors such as macroeconomic issues, external shocks, and fashion trends \citep{chen2019dynamic, keskin2022selling}. Therefore, to incorporate high variance demand patterns and demand evolvement in the marketplace, it is necessary to consider correlation between customer arrivals.

In this paper, we study the NRM problem with correlated customer arrivals and aim to answer the following two research questions. First, how should we model correlated customer arrivals in the NRM problem? Existing literature models the high variance demand pattern and demand evolvement pattern separately. We propose a unified model that incorporates both patterns in a single framework. Second, as the optimal policy is computationally intractable due to the curse of dimensionality, how can we design a near-optimal policy with strong theoretical guarantees? A key step in designing a near-optimal policy is to find a sound approximation of the optimal policy. We propose an approximation under the unified correlated customer arrival model and derive our policy. We measure the performance of our policy by the approximation ratio, which is defined as the relative difference between the expected total reward collected by our policy and that of the optimal policy. The formal definition of the approximation ratio is provided in \Cref{sec:problem} after introducing the notations and problem formulation.

\subsection{Main Results and Contributions}

Our main results can be summarized into three parts. First, we propose a new model of correlated customer arrivals that unifies previous models. Our model assumes the existence of a system state that transits according to a time-inhomogeneous Markov chain. This allows us to capture previous correlation models while providing greater flexibility in modeling customer arrival patterns. Second, we present a new approximation of the optimal policy that serves as an upper bound on the expected total reward collected by the optimal policy. We demonstrate that our upper bound is asymptotically optimal compared to the optimal policy, as the initial capacities scaled up to infinity. Third, we derive a near-optimal policy with an approximation ratio of $1/(1+L)$, where $L$ denotes the maximum number of resources that a customer would require. Our policy specifies bid prices for each system state and time period, and we serve the customer only if the reward exceeds the associated bid prices. In this way, our policy can be viewed as a generalization of the classical bid price control policy for the NRM problem (e.g. \cite{talluri1998analysis}, \cite{adelman2007dynamic}) to the correlated arrival setting. We now provide further illustrations of our main results and make comparisons with previous results in detail.


\subsubsection{A unified model of correlated customer arrivals.} To formalize the concept of having a system state and time-inhomogeneous transition probabilities, we denote the state of the system at each period $t$ as $\bs_t$, which synthesizes the current system information. For example, in settings with finite types of customers, we can let $\bs_t$ represent the type $j_t$ of customer $t$, where each type refers to a certain subset of resources required by the customer and a corresponding reward. The meaning of the state can also extend beyond customer type. In inventory management literature (e.g., \cite{song1993inventory, sethi1997optimality, chen2001optimal}), the system state refers to the quantity of demand, while in revenue management literature, the system state can represent other statistical characterizations of customers (e.g., \cite{aviv2005partially}). The system transits to a new state according to a given probability, denoted by $p_t(\bs_t=\bs, \bs_{t+1}=\bs')$ for any possible states $\bs$ and $\bs'$ in the state space. The transition of the system state corresponds to the transition on a time-inhomogeneous Markov chain. Therefore, we name our arrival model the Markovian-correlated arrival model. In contrast to the Markovian-modulated arrival model studied in previous literature, we allow the transition probabilities $p_t(\bs_t=\bs, \bs_{t+1}=\bs')$ to vary across periods. This generalization allows the total number of arrivals of customers to have an arbitrary distribution that allows the coexistence of large demand volume and large demand variability. We illustrate this point in details below.

In order to show that the distribution of the total number of customer arrivals can have large volume and large variability simultaneously, we let a random variable $D$ capture the total number of customer arrivals, which can follow any distribution. Then, a survival rate $\rho_t=P(D\geq t+1| D\geq t)$ is defined.
At each period $t$, given $D\geq t$, one customer arrives, and customer $t$ is of type $j\in[n]$ with probability $\lambda_{j,t}$. Our Markovian-correlated model can capture the model described above by letting the state space be $\{0, 1, \dots, n\}$, where state $j\in[n]$ denotes that a customer of type $j$ has arrived, and state $0$ denotes no customer arrival. We define the transition probabilities in our model as follows:
\[
p_t(j, j')=\rho_{t-1}\cdot\lambda_{j',t},~ p_t(j,0)=1-\rho_{t-1},\text{~and~}p_t(0,0)=1, p_t(0,j)=0~\forall j, j'\in[n], \forall t\in[T].
\]
It is easy to see that the distribution of the total number of customer arrivals is equivalent to the distribution of $D$, which can be any distribution. 

As illustrated above, our Markovian-correlated model not only captures the demand evolvement but also allow the total number of customer arrivals to have an arbitrary distribution. In \Cref{sec:Corremodel}, we provide detailed comparisons with other correlated arrival models proposed in previous literature. Overall, our model provides a more flexible and accurate way to capture customer arrival patterns in many business and economic settings.

\subsubsection{A new LP upper bound of the optimal policy.} To derive the optimal policy, we typically solve the dynamic programming (DP) problem. However, due to the curse of dimensionality, solving the DP becomes computationally intractable for large-scale systems. Therefore, it is common to derive a sound approximation of the optimal policy that is computationally tractable. This approximation not only enables practical policy implementation but also serves as an upper bound for the expected total reward collected by the optimal policy, proving the performance guarantee of our policy. For independent customer arrival models, the most prevalent approximation is the fluid approximation (e.g., \cite{gallego1994optimal}). The fluid approximation is derived by computing the expected consumption of resource capacities in the constraints and the total expected revenue in the objective function through customer arrival distributions. For independent customer arrivals, the fluid approximation is asymptotically tight as the system size scales up \citep{gallego1994optimal}. Constant approximation ratios have also been obtained for the fluid approximation under various settings (e.g., \cite{ma2020approximation, baek2022bifurcating}). However, when customer arrivals are correlated, the performance of the optimal policy can be arbitrarily worse than that of the fluid approximation, as shown in \cite{bai2022fluid} and \cite{aouad2022nonparametric}. Therefore, it is essential to propose new approximations for optimal policy with correlated customer arrivals.

Although the DP is computationally intractable, we exploit the DP formulation to derive our approximation. Motivated by the broad literature on approximate DP (e.g., \cite{powell2007approximate}), we use a linear combination over the remaining resources to approximate the value-to-go function. Specifically, we denote $V^*_t(\bc, \bs)$ as the value-to-go function of the DP at period $t$, given the remaining capacities $\bc=(c_1,\dots, c_m)$ and the current state $\bs$. We approximate $V^*_t(\bc, \bs)$ by the following linear formulation:
\[
V^*_t(\bc, \bs) \approx \theta^t(\bs)+\sum_{i\in[m]}c_i\cdot \beta_{i}^t(\bs),
\]
where $\theta^t(\bs)$ and $\{\beta^t_i(\bs), \forall i\in[m]\}$ are weights determined for each period $t$ and each state $\bs$. We develop a linear programming (LP) model for computing the weights $\{\theta^t(\bs), \beta^t_i(\bs), \forall i\in[m], \forall t, \forall \bs\}$ and show that its optimal objective value is an upper bound on the expected total reward collected by the optimal policy (the DP value). We also demonstrate that our upper bound is asymptotically optimal compared to the optimal policy as the initial capacities scales up to infinity, as discussed in \Cref{sec:OtherUB}. Overall, our proposed LP well approximates the optimal policy and provides an efficient way to derive the policy for settings with correlated customer arrivals.

\subsubsection{A bid price control policy and performance guarantee.} Motivated by backward induction in the DP formulation, we develop a variant of the bid price control policy. Specifically, for each customer type $j$, given the period $t$ and the system state $\bs$, we assign a bid price $\nu^{t}_j(\bs)$. We then use a linear combination of the bid prices to represent the marginal benefits of having an extra $\ba_{j_t}$ units of resources, and a customer is served only if its reward exceeds the marginal benefits. The bid prices are computed in a backward induction manner that mimics the backward induction in the approximate DP formulation. We also show that the constructed bid prices can be converted into a feasible solution for our LP upper bound of the optimal policy. By following these steps, we show that our policy enjoys an approximation ratio bound of $1/(1+L)$, where $L$ denotes the maximum number of resources that a customer will consume. Our approximation ratio bound guarantees that our policy performs reasonably well compared to the optimal policy, irrespective of the number of time periods, resources, and customer types.
Finally, we extend all our results to the more general setting with customer choices, where the decision maker offers an assortment at each period, and the customer chooses one product from the assortment. As detailed in \Cref{sec:Choicemodel}, our extension to the more general setting with customer choices enhances the practicality and applicability of our results.


\subsection{Other Related Literature}

We now review other related literature. We first discuss previous methods in the literature on correlated customer arrivals and we then compare against the constrained Markov decision process literature. We finally discuss existing literature on developing near-optimal policies for NRM problems.

The Markovian-modulated demand process is a prevalent way to model correlated customer arrivals, where the system state transits according to a Markov chain over time. This process has been widely studied in the literature of inventory management and supply chain management, and interested readers can refer to \cite{simchi2004handbook} for an overview. The Markovian-modulated demand process has also been extensively studied in the pricing literature. For example, \cite{rustichini1995learning} assumes the system state evolves according to a two-state Markov chain, while \cite{aviv2005partially} considers a partially observed Markovian-modulated setting where the system state is unobserved, and only the demand can be observed. \cite{feng2000perishable} considers a revenue management problem where the arrivals of the customers depend on the sales to date, which can be captured in a Markovian way. The pricing problem for Markovian-modulated settings has also been considered in \cite{den2018dynamic} under the context of inventory control. \cite{keskin2022selling} assumes the transition probabilities to be unknown and considers pricing with learning. Moreover, in \cite{jia2023online}, the Markovian-modulated setting is considered for an online resource allocation problem with a single resource and time-homogeneous transition probabilities. 


Notably, by assuming a Markovian-correlated customer arrival model, our problem falls into the broad literature of constrained Markov decision processes (CMDP), where there are resource constraints over the entire horizon, and the action taken at each period consumes certain resources. Various methods have been proposed to solve CMDP near-optimally, including a linear programming-based approach and a Lagrangian approach (see \cite{altman1999constrained} and the references therein). The reinforcement learning counterpart of CMDP has also been studied in the literature, where the transition probabilities are assumed to be unknown. The problem is studied in various settings, including \cite{wei2018online, qiu2020upper} for adversarial objectives, \cite{zheng2020constrained} for bandit feedback, and \cite{efroni2020exploration, germano2023best} for more general settings. However, the methods developed in these papers are for the infinite horizon (discounted) average reward setting. For finite horizon settings, only sublinear regret bounds are proved. Compared to the literature above, we are the first to achieve a constant approximation ratio bound for a CMDP with an NRM formulation, where our bound holds irrespective of the number of time periods. 
Our results extend the existing literature and provide a new framework for solving CMDP problems with NRM formulations in finite horizon settings.

The optimal policy for the NRM problem can be characterized by DP. However, due to the curse of dimensionality, the DP is computationally intractable. Therefore, one mainstream of literature on the NRM problem is to develop approaches that approximate the DP solution. In \cite{adelman2007dynamic}, the author proposes using a linear combination over the remaining capacities to approximate the value-to-go function, and the coefficients in the linear combination can be computed through solving a LP, which gives a tighter upper bound than the traditional fluid approximation. Subsequently, the approximate DP approach is further investigated in the literature (e.g., \cite{zhang2009approximate, farias2007approximate, zhang2011improved, meissner2012network, tong2014approximate, kunnumkal2016piecewise, zhang2022product}) under various settings for the NRM problem. Notably, \cite{ma2020approximation} develops a non-linear approximation to the DP value and obtains a constant approximation ratio. \cite{baek2022bifurcating} and \cite{simchi2022online} further investigate the reusable resource settings. Compared to the previous literature on approximate DP approach for NRM problem, we consider correlated customer arrivals. The Lagrangian relaxation approach has also been investigated in the literature for approximating the DP value. For example, \cite{topaloglu2009using} applies the Lagrangian relaxation approach to the NRM problem, and \cite{brown2014information} applies it to general DP. However, these previous methods do not enjoy a constant approximation ratio bound.

In addition to the approximation ratio bound, the revenue loss bound has also been widely studied for the NRM problem, which measures the additive difference between the expected total reward collected by the proposed policy and that of the optimal policy. One popular way to derive a policy with a strong revenue loss bound is to consider the fluid approximation and use its optimal solution to derive the policies. For example, \cite{talluri1998analysis} proposes a static bid-price policy based on the dual variable of the fluid approximation and proves that the revenue loss is $O(\sqrt{T})$. Subsequently, \cite{reiman2008asymptotically} shows that by re-solving the fluid approximation once, one can obtain an $o(\sqrt{T})$ upper bound on the revenue loss. Then, \cite{jasin2012re} shows that under a non-degeneracy condition for the fluid approximation, a policy that re-solves the fluid approximation at each time period will lead to an $O(1)$ revenue loss, which is independent of the horizon $T$. A later paper \cite{jasin2013analysis} further discusses the relationship between the performances of the control policies and the number of times of re-solving the fluid approximation. Recently, by considering a tighter prophet upper bound, \cite{bumpensanti2020re} propose an infrequent re-solving policy and show that their policy achieves an $O(1)$ upper bound on the revenue loss even without the "non-degeneracy" assumption. With a different approach, \cite{vera2021bayesian} proves the same $O(1)$ upper bound for the NRM problem. Their approach is further generalized in \cite{vera2021online, freund2019good, freund2021overbooking} for other online decision-making problems. When there can be an infinite number of customer types, a logarithmic revenue loss is achieved in \cite{balseiro2021survey}, \cite{besbes2022multisecretary}, \cite{bray2022logarithmic}, and \cite{jiang2022degeneracy} under various conditions. For price-based NRM problems, an $O(1)$ revenue loss is proved in \cite{wang2022constant} for the resolving heuristics. Notably, \cite{bai2022fluid} and \cite{li2023revenue} derive policies that are asymptotically optimal under various correlated arrival models, which translate into a sublinear regret. Overall, the literature on the revenue loss bound for the NRM problem is extensive and covers various settings and approaches. These results provide valuable insights into the performance of different policies and help practitioners design more effective and efficient revenue management strategies. Our work complements the literature on NRM problem by providing a constant approximation ratio with correlated customer arrivals.


\section{Problem Formulation}\label{sec:problem}
We consider a network revenue management problem, where there are $m$ resources and each resource $i\in[m]$ has an initial capacity $C_i\in\mathbb{R}_{\ge0}$. There are $n$ products and we have a binary variable $a_{i,j}\in\{0,1\}$ denoting whether product $j$ would require one unit of resource $i$ to produce, for each $j\in[n]$ and $i\in[m]$.
There are $T$ discrete time periods and at each period $t\in[T]$, one customer arrives, denoted as customer $t$. Customer $t$ would require one product and we call customer $t$ a \textit{type-$j$} customer if customer $t$ requires product $j$. Each customer is associated with a size and a reward. For a customer of type $j$, the size is denoted by $\ba_j=(a_{j,1},\dots,a_{j,m})\in\{0,1\}^m$ and the reward is denoted by $r_j$. We assume that the type of customer $t$ is realized as $j$ with a given probability, for each $j\in[m]$. However, this probability would depend on the type realizations of previous customers, which reflects the correlation of customer arrivals. To be more concrete, we use $\bs_t$ to denote the \textit{state} of period $t$ and we denote by $\mathcal{S}$ the state space. The type of customer $t$ is determined by the system state. Given $\bs_t$, the type of customer $t$ is determined as $j(\bs_t)$.

After query $t$ arrives and the state $\bs_t$ is revealed, the decision maker has to decide immediately and irrevocably whether or not to serve customer $t$. Note that customer $t$ can only be served if for every resource $i$ its remaining capacity is at least $a_{j(\bs_t), i}$. By serving customer $t$, each resource $i$ will be consumed by $a_{j(\bs_t), i}$ units and a reward $r_{j(\bs_t)}$ will be collected by the decision maker. Then, in the next period $t+1$, the state would transfer to $\bs_{t+1}$ with probability $p_t(\bs_t, \bs_{t+1})$.
The goal of the decision maker is to maximize the total reward collected during the entire horizon subject to the resource capacity constraints.

Any online policy $\pi$ for the decision maker is specified by a set of decision variables $\{x_{t}^{\pi}\}_{\forall t\in[T]}$, where $x_t^{\pi}$ is a binary variable and denotes whether customer $t$ is served or not, for all $t\in[T]$. Note that $x_t^{\pi}$ can be stochastic if $\pi$ is a randomized policy. Any policy $\pi$ is feasible if for all $t\in[T]$, $x_t^{\pi}$ depends only on the problem instance $\left\{p_t(\bs, \bs'), \forall t\in[T], \forall \bs, \bs'\in\mathcal{S}\right\}$ and the realizations up to now $\{\bs_1, \dots, \bs_t\}$, and the following capacity constraint is satisfied:
\begin{equation}\label{eqn:capacityconstraint}
\sum_{t=1}^T a_{j(\bs_t),i}\cdot x_t^{\pi}\leq C_i,~~\forall i\in[m].
\end{equation}
The total collected value of policy $\pi$ is given by $V^{\pi}(I)=\sum_{t=1}^T r_{j(\bs_t)}\cdot x_t^{\pi}$, where $I=\{\bs_t, \forall t\in[T]\}$ denotes the sample path.

Our goal is to develop a feasible polynomial-time online policy $\pi$ for the decision maker to maximize $\mathbb{E}_{I\sim \mathcal{F}}[V^{\pi}(I)]$, where we use $\mathcal{F}=\left\{p_t(\bs, \bs'), \forall t\in[T], \forall \bs, \bs'\in\mathcal{S}\right\}$ to denote the problem instance for notational simplicity. The benchmark is the \textit{optimal online policy}, which we denote by $\pi^*$ and can be obtained as the solution to the following problem:
\begin{equation}\label{eqn:optimal}
    \pi^*=\text{argmax}_{\pi} \mathbb{E}_{I\sim \mathcal{F}}[V^{\pi}(I)]
\end{equation}
For any feasible online policy $\pi$, we use \textit{approximation ratio} to measure its performance, which is defined as follows:
\begin{equation}\label{eqn:regret}
    \gamma(\pi):= \inf_{\mathcal{F}} \frac{\mathbb{E}_{I\sim \mathcal{F}}[V^{\pi}(I)]}{\mathbb{E}_{ I\sim \mathcal{F}}[V^{\pi^*}(I)]}.
\end{equation}
Note that in the definition \eqref{eqn:regret}, the performance of the online policy $\pi$ is minimized over the problem instance given by the probabilities $\left\{p_t(\bs, \bs'), \forall t\in[T], \forall \bs, \bs'\in\mathcal{S}\right\}$, instead of the support for customers' size and reward $\{(r_j, \ba_j), \forall j\in[n]\}$. Indeed, as we will show later (see discussions in \Cref{sec:proofCR}), the approximation ratio of any feasible online policy would inevitably depend on the support $\{(r_j, \ba_j), \forall j\in[n]\}$. As a result, the approximation ratio of our policy also depends on $\{(r_j, \ba_j), \forall j\in[n]\}$. We compare our arrival model with other arrival models in the literature in \Cref{sec:Corremodel}.

\section{An Approximation of the Optimal Policy}\label{sec:UB}
In this section, we derive an approximation of the optimal policy to serve as an upper bound of the expected total reward collected by the optimal online policy. The upper bound is derived from the dynamic programming (DP) formulation of the optimal online policy, where we would regard the combination of the remaining capacities of each resource, denoted by $\bc_t=(c_{t,1},\dots, c_{t,m})$, and $\bs_t$ as the current state in the DP formulation. However, due to the curse of dimensionality, the state space for the DP can be exponentially large. Therefore, we apply a linear approximation to the DP formulation, which not only enables us to derive a linear programming (LP) to serve as an upper bound to the optimal online policy but also implies our online policy.

Denote by $V_{t}^*(\bc, \bs)$ the value to go function at period $t$, given the remaining capacity $\bc$ and the current state $\bs$. The backward induction can be given as follows:
\begin{equation}\label{eqn:backward}
\begin{aligned}
V^*_{t}(\bc, \bs)=&\max\left\{ \bI_{\bc\geq\ba_{j(\bs)}}\cdot\left( r_{j(\bs)}+\sum_{\bs'\in\mathcal{S}}p_t(\bs, \bs')\cdot V^*_{t+1}(\bc-\ba_{j(\bs)}, \bs')  \right), \sum_{\bs'\in\mathcal{S}}p_t(\bs, \bs')\cdot V^*_{t+1}(\bc, \bs')
 \right\}.
\end{aligned}
\end{equation}
Then, the expected collected reward for the optimal online policy can be given by the DP value, i.e., $\mathbb{E}_{I\sim \mathcal{F}}[V^{\pi^*}(I)]=\sum_{\bs\in\mathcal{S}}p_1(\bs)\cdot V^*_{1}(\bm{C}, \bs)$ where we use $p_1(\bs)$ to denote the probability that the initial state $\bs_1$ is realized as $\bs$, for all $\bs\in\mathcal{S}$. Note that the backward induction \eqref{eqn:backward} admits the following equivalent LP formulation:
\begin{subequations}\label{lp:DP}
\begin{align}
\min ~~& \sum_{\bs\in\mathcal{S}}p_1(\bs)\cdot V_{1}(\bm{C}, \bs) \\
\mbox{s.t.} ~~& V_{t}(\bc,\bs)\geq r_{j(\bs)}+\sum_{\bs'\in\mathcal{S}}p_t(\bs, \bs')\cdot V_{t+1}(\bc-\ba_{j(\bs)}, \bs'), \forall \bc\geq\ba_{j(\bs)}, \forall t\in[T], \forall \bs\in\mathcal{S} \label{const:DP1}\\
& V_{t}(\bc,\bs)\geq \sum_{\bs'\in\mathcal{S}}p_t(\bs, \bs')\cdot V_{t+1}(\bc, \bs'), \forall \bc, \forall t\in[T], \forall \bs\in\mathcal{S} \label{const:DP2}\\
& V_{t}(\bc,\bs)\geq 0, \forall \bc, \forall t\in[T], \forall \bs\in\mathcal{S}.
\end{align}
\end{subequations}
Here, we regard each $V_{t}(\bc, \bs)$ as a decision variable, which represents the dynamic programming value-to-go $V^*_{t}(\bc, \bs)$. It is known that in an optimal solution, the decision variable $V_{t}(\bc, \bs)$ equals $V^*_{t}(\bc, \bs)$ (e.g. \cite{schweitzer1985generalized, de2003linear, adelman2007dynamic}). However, since there can be exponentially many possible values for $\bc$, the LP \eqref{lp:DP} has exponentially many decision variables. We apply a linear approximation to reduce the number of decision variables. To be specific, we restrict the formulation of the decision variable $V_{t}(\bc, \bs)$ into the following linear formulation:
\begin{equation}\label{eqn:approximate}
V_{t}(\bc, \bs)=\theta^t(\bs)+\sum_{i\in[m]}c_i\cdot \beta_{i}^t(\bs),~~\forall \bc
\end{equation}
where $\left\{\theta^t(\bs), \beta^t_{i}(\bs), \forall i\in[m], \forall \bs\in\mathcal{S}, \forall t\in[T] \right\}$ is a set of \textit{non-negative} parameters to be determined later. Plugging the linear approximation \eqref{eqn:approximate} into the LP \eqref{lp:DP}, we derive a simplified (and further relaxed) LP with decision variables being $\left\{\theta^t(\bs), \beta^t_{i}(\bs), \forall i\in[m], \forall \bs\in\mathcal{S}, \forall t\in[T] \right\}$.
\begin{subequations}\label{lp:ADP}
\begin{align}
\min ~~&\sum_{\bs\in\mathcal{S}} p_1(\bs)\cdot \left( \theta^1(\bs)+\sum_{i\in[m]}C_i\cdot\beta^1_{i}(\bs) \right)\label{Ob:ADP}\\
\mbox{s.t.}~~& \theta^t(\bs)-\sum_{\bs'\in\mathcal{S}}p_t(\bs, \bs')\cdot\theta^{t+1}(\bs')\geq \left[ r_{j(\bs)}-\sum_{\bs'\in\mathcal{S}}p_t(\bs, \bs')\cdot\sum_{i\in[m]}a_{i,j(\bs)}\cdot \beta^{t+1}_{i}(\bs') \right]^+  \nonumber\\
&~~~~~~~~+\sum_{i\in[m]}C_i\cdot\left[ \sum_{\bs'\in\mathcal{S}}p_t(\bs, \bs')\cdot \beta^{t+1}_{i}(\bs')-\beta_{i}^t(\bs) \right]^+, ~~~~~\forall t\in[T], \forall \bs\in\mathcal{S}  \label{const:ADP}\\
&\theta^t(\bs)\geq0, \beta^t_{i}(\bs)\geq0, ~~~~~\forall i\in[m], \forall t\in[T], \forall \bs\in\mathcal{S}.
\end{align}
\end{subequations}
Note that though there are $[\cdot]^+$ operators, the above optimization problem is indeed equivalent to an LP. To see this point, for each $[\cdot]^+$ operator, we can introduce a new non-negative variable to represent its value. For example, for each $t\in[T]$ and $\bs\in\mathcal{S}$, we can introduce a new decision variable $\eta_t(\bs)$ with new constraints
\[
\eta_t(\bs)\geq0 \text{~and~} \eta_t(\bs)\geq \sum_{\bs'\in\mathcal{S}}p_t(\bs, \bs')\cdot \beta^{t+1}_{i}(\bs')-\beta_{i}^t(\bs).
\]
It is clear to see that in an optimal solution, $\eta_t(\bs)$ would represent the value of $\left[ \sum_{\bs'\in\mathcal{S}}p_t(\bs, \bs')\cdot \beta^{t+1}_{i}(\bs')-\beta_{i}^t(\bs) \right]^+$ in the constraint \eqref{const:ADP}. The formulation \eqref{lp:ADP} is chosen to simplify the derivation as we will construct a feasible solution to \eqref{lp:ADP} to prove our approximation ratio bound, as detailed in \Cref{sec:Construct}.

In the following lemma, we show that the optimal objective value of LP \eqref{lp:ADP} serves as an upper bound of the optimal online policy, with the formal proof relegated to \Cref{sec:pf3}.
\begin{lemma}\label{lem:LPupperbound}
Denote by $\hat{V}^*$ the optimal objective value of LP \eqref{lp:ADP}. It holds that
\[
\hat{V}^*\geq \mathbb{E}_{I\sim \mathcal{F}}[V^{\pi^*}(I)].
\]
\end{lemma}
Therefore, throughout the paper, we compare against LP \eqref{lp:ADP}.

\subsection{Discussion on the Asympototic Optimality}\label{sec:OtherUB}

We now show the asymptotic optimality our upper bound LP \eqref{lp:ADP} with respect to the optimal policy, as the initial capacities scale up to infinity. We illustrate this asymptotic optimality under an important arrival model, the high-variance correlated model studied in \cite{bai2022fluid} and \cite{aouad2022nonparametric}, which can be captured in our Markovian model as follows. The high-variance model can be described as having the state $\mathcal{S}=\{0,1,\dots, n\}$, where state $j\in[m]$ denotes the customer is of type $j$ and the state $0$ denotes there is no customer arrival. Also, for each period $t\in[T]$, conditional on customer $t$ arrives, customer $t$ is of type $j\in[n]$ with probability $\lambda_{j,t}$. Note that under the high-variance correlated model, it has been shown that the traditional fluid upper bound that is widely used in the previous literature on independent customer arrivals (e.g. \cite{gallego1994optimal}) can be arbitrarily bad compared to the optimal policy. Therefore, new LP upper bounds are developed to approximate the optimal policy and the asymptotic optimalities of the upper bounds are established. We now show that our upper bound LP \eqref{lp:ADP} is also asymptotic optimal with respect to the optimal policy. The formal proof of \Cref{prop:tightUB} is relegated to \Cref{sec:pf3}.

\begin{proposition}\label{prop:tightUB}
Under the high-variance correlated model, the upper bound LP \eqref{lp:ADP} is asymptotically optimal with respect to the optimal policy in that
\[
\frac{\mathbb{E}_{ I\sim \mathcal{F}}[V^{\pi^*}(I)]}{\text{LP~}\eqref{lp:ADP}}\geq 1-O(1/\sqrt{\min_{i\in[m]}C_i}).
\]
with $\mathbb{E}_{ I\sim \mathcal{F}}[V^{\pi^*}(I)]$ denoting the expected total reward collected by the optimal online policy.
\end{proposition}

\section{Description of Our Policy}\label{sec:Policy}

In this section, we derive our policy. Our policy is motivated by the DP formulation given in \eqref{eqn:backward}. Note that in the DP, given the state $\bs_t$, we serve customer $t$ as long as $\bc_t\geq\ba_{j(\bs_t)}$ and
\begin{equation}\label{eqn:042601}
r_{j(\bs_t)}\geq \sum_{\bs'\in\mathcal{S}}p_t(\bs_t, \bs')\cdot \left( V^*_{t+1}(\bc_t, \bs')-V^*_{t+1}(\bc_t-\ba_{j(\bs_t)}, \bs') \right),
\end{equation}
which follows directly from the DP formulation \eqref{eqn:backward}. We follow the same intuition of the decision rule in \eqref{eqn:042601} to derive our policy. To be specific, denote by $\pi$ our policy and denote by $H^{\pi}_{t}(\bc, \bs)$ the total expected reward collected by the policy $\pi$ from period $t$ to period $T$, given the remaining capacity at period $t$ is $\bc$ and the state of period $t$ is $\bs$. We set $y^{\pi}_t(\bs_t)=1$ and serve customer $t$ as long as $\bc_t\geq\ba_{j(\bs_t)}$ and
\begin{equation}\label{eqn:042602}
r_{j(\bs_t)}\geq \sum_{\bs'\in\mathcal{S}}p_t(\bs_t, \bs')\cdot \left( H^{\pi}_{t+1}(\bc_t, \bs')-H^{\pi}_{t+1}(\bc_t-\ba_{j(\bs_t)}, \bs') \right).
\end{equation}
Here, the term $H^{\pi}_{t+1}(\bc_t, \bs')-H^{\pi}_{t+1}(\bc_t-\ba_{j(\bs_t)}, \bs')$ denotes the marginal increase in terms of the total expected reward collected by $\pi$ if we do not serve customer $t$, given $\bs_{t+1}$ is realized as $\bs'$. We compute the marginal increase for not serving customer $t$ by taking an expectation over $\bs_{t+1}$. Customer $t$ is served in a myopic sense if the benefits of serving customer $t$, which is $r_{j(\bs_t)}$, exceeds the expected marginal increase for not serving customer $t$.

The only issue of implementing the decision rule \eqref{eqn:042602} relies on how to compute the term $H^{\pi}_{t+1}(\bc_t, \bs')-H^{\pi}_{t+1}(\bc_t-\ba_{j(\bs_t)}, \bs')$. One could indeed use backward induction and follow the decision rule \eqref{eqn:042602} to compute the value of $H^{\pi}_{t+1}(\bc_t, \bs')$ for every possible $\bc_t$ and every $\bs'\in\mathcal{S}$. However, the computational complexity would be the same as directly solving the DP and we will encounter the curse of dimensionality, i.e., the computational complexity would be exponential in $m$. In order to remedy this issue, we follow the idea of bid price control (e.g. \cite{talluri1998analysis}, \cite{adelman2007dynamic}) and derive a set of bid prices to approximate the term $H^{\pi}_{t+1}(\bc_t, \bs')-H^{\pi}_{t+1}(\bc_t-\ba_{j(\bs_t)}, \bs')$.

We introduce a bid price $\nu^{t}_j(\bs')$ for each $t\in[T]$, each $j\in[n]$ and each $\bs'\in\mathcal{S}$. Note that each type of customer can require multiple resources simultaneously to be served. For any type $j\in[n]$, we denote by $\mathcal{A}_j$ the set of resources that will be consumed by serving type $j$ customer, i.e., $\mathcal{A}_j=\{ \forall i\in[m]: a_{i,j}=1 \}$. Analogously, for any resource $i\in[m]$, we denote by $\mathcal{B}_i$ the set of customer types that would require one unit of resource $i$ to be served, i.e., $\mathcal{B}_i=\{ \forall j\in[n]: a_{i,j}=1 \}$. We use the bid price $\nu^{t}_j(\bs')$ to approximate the marginal gain we can obtain from serving customers of type $j$ if we have \textit{one more unit of each resource $i\in\mathcal{A}_j$}, no matter what remaining resources are. Then, the benefits of having one more resource $i$ at period $t$ given state $\bs'$ can be approximated by $\frac{1}{C_i}\cdot\sum_{j\in\mathcal{B}_i} \nu^t_j(\bs')$, where we normalize by the total capacity $C_i$ to guarantee that the approximation is valid no matter what remaining resources are.
Comparing $\bc_t$ and $\bc_t-\ba_{j(\bs_t)}$, we have one more unit of resource $i$ for each $i\in\mathcal{A}_{j(\bs)}$. Therefore, the term $H^{\pi}_{t+1}(\bc_t, \bs')-H^{\pi}_{t+1}(\bc_t-\ba_{j(\bs_t)}, \bs')$ can be approximated by
\[
\sum_{i\in\mathcal{A}_{j(\bs_t)}}\frac{1}{C_i}\cdot\sum_{j'\in\mathcal{B}_i}\nu^{t+1}_{j'}(\bs').
\]
The decision rule \eqref{eqn:042602} can be finally represented as serving customer $t$ as long as $\bc_t\geq\ba_{j(\bs)}$ and
\begin{equation}\label{eqn:042603}
r_{j(\bs_t)}\geq \sum_{\bs'\in\mathcal{S}}p_t(\bs_t, \bs')\cdot\sum_{i\in\mathcal{A}_{j(\bs_t)}}\frac{1}{C_i}\cdot\sum_{j'\in\mathcal{B}_i}\nu^{t+1}_{j'}(\bs').
\end{equation}
Our bid price policy is formally described in \Cref{alg:BMP}, where we approximate the expected marginal increase $\sum_{\bs'\in\mathcal{S}}p_t(\bs_t, \bs')\cdot \left( H^{\pi}_{t+1}(\bc_t, \bs')-H^{\pi}_{t+1}(\bc_t-\ba_{j(\bs_t)}, \bs') \right)$ and we serve customer $t$ when the reward $r_{j(\bs_t)}$ exceeds the approximation of the expected marginal increase.

\subsection{Bid Price Computing}\label{sec:Bidcompute}

We now describe how to compute the bid price $\nu^t_{j}(\bs)$ for all $j\in[n]$, $t\in[T]$ and all $\bs\in\mathcal{S}$. 
The bid price $\nu^t_{j}(\bs)$ is computed in a backward induction from $T$ to $1$. To be specific, we set $\nu^{T+1}_{j}(\bs)=0$ for any $j\in[n]$ and any $\bs\in\mathcal{S}$. Then, for $t=T, T-1,\dots, 1$, we compute iteratively
\begin{equation}\label{eqn:042202}
    \nu^t_j(\bs)=\sum_{\bs'\in\mathcal{S}}p_t(\bs, \bs')\cdot\nu^{t+1}_j(\bs')+\bI_{\{j=j(\bs)\}} \cdot\left[r_{j}-\sum_{\bs'\in\mathcal{S}}p_t(\bs, \bs')\cdot\sum_{i\in\mathcal{A}_j}\frac{1}{C_i}\cdot\sum_{j'\in\mathcal{B}_i}\nu^{t+1}_{j'}(\bs') \right]^+
\end{equation}
for every $j\in[n]$ and every $\bs\in\mathcal{S}$. We provide an interpretation of the intuition behind the backward induction \eqref{eqn:042202}. As we have illustrated previously, the term $\sum_{\bs'\in\mathcal{S}}p_t(\bs, \bs')\cdot\sum_{i\in\mathcal{A}_j}\frac{1}{C_i}\cdot\sum_{j'\in\mathcal{B}_i}\nu^{t+1}_{j'}(\bs')$ is an approximation of the expectation of the marginal increase for having an extra $\ba_{j}$ resources. Then, the term $\sum_{\bs'\in\mathcal{S}}p_t(\bs, \bs')\cdot\sum_{i\in\mathcal{A}_j}\frac{1}{C_i}\cdot\sum_{j'\in\mathcal{B}_i}\nu^{t+1}_{j'}(\bs')$ can be interpreted as the ``opportunity cost'' for serving customer $t$ with type $j$. As a result, by noting the decision rule \eqref{eqn:042603}, the term
\[
\left[r_{j}-\sum_{\bs'\in\mathcal{S}}p_t(\bs, \bs')\cdot\sum_{i\in\mathcal{A}_j}\frac{1}{C_i}\cdot\sum_{j'\in\mathcal{B}_i}\nu^{t+1}_{j'}(\bs')\right]^+
\]
accounts for the gain of our policy for the period $t$, which will be used to update the bid price $\nu^t_j(\bs)$. The bid price update \eqref{eqn:042202} also corresponds to our decision rule \eqref{eqn:042603}. Note that we only need to update the bid price $\nu^t_j(\bs)$ for $j=j(\bs)$ since the type $j(\bs)$ customer arrives at period $t$ given the state $\bs$.

The backward induction \eqref{eqn:042202} can also be motivated by the approximate DP approach from \cite{ma2020approximation} with the differences described as follows. For each period $t$, we define the bid price $\nu^t_j(\bs)$ for each type $j\in[n]$ and each state $\bs\in\mathcal{S}$ while their approximate DP parameters are only defined for each type $j\in[n]$. We let our bid price depend on the state to deal with the correlation of the customer arrivals. Also, for each state $\bs$, we update the bid price $\nu^t_j(\bs)$ if and only if $j=j(\bs)$, while their approximate DP parameters are updated for each $j\in[n]$. Finally, we need to take expectations over the state of the next period given the current state in our update \eqref{eqn:042202}, which is not needed in the approximate DP approach since they assumed independent customer arrival of each period.


\begin{algorithm}[ht!]
\caption{Bid Price Policy}
\label{alg:BMP}
\begin{algorithmic}[1]
\State Compute the bid prices $\nu^t_j(\bs)$ for any $j\in[n]$, any $t\in[T]$ and any $\bs\in\mathcal{S}$, following the backward induction equation described in \eqref{eqn:042202}.
\For{t=1,\dots,T}
\State observe the state $\bs_t$ and the remaining capacities $\bc_t$.
 \If {there exists a resource $i\in[m]$ such that $c_{t,i}< a_{i, j(\bs)}$} {reject customer $t$}
\Else {~serve customer $t$ if and only if \begin{equation}\label{eqn:042904} r_{j(\bs)}\geq \sum_{\bs'\in\mathcal{S}}p_t(\bs, \bs')\cdot\sum_{i\in\mathcal{A}_{j(\bs)}}\frac{1}{C_i}\cdot\sum_{j'\in\mathcal{B}_i}\nu^{t+1}_{j'}(\bs').\end{equation}}
\EndIf
\EndFor
\end{algorithmic}
\end{algorithm}

\section{Algorithm Analysis}\label{sec:Analysis}

In this section, we prove the approximation ratio bound of our \Cref{alg:BMP}. Specifically, denote by
\begin{equation}\label{eqn:042203}
L=\max_{j\in[n]}\left\{
\sum_{i\in[m]}a_{i,j} \right\}.
\end{equation}
We show that \Cref{alg:BMP} would collect an expected total reward at least $1/(1+L)$ fraction of that of the optimal policy. Our proof is based on our construction of the bid price $\nu^t_j(\bs)$ described in \eqref{eqn:042202} and can be classified into two steps. In the first step, we show that the expected total reward collected by \Cref{alg:BMP} can be lower bounded by $\sum_{\bs\in\mathcal{S}}p_1(\bs)\cdot\sum_{j\in[n]}\nu^1_j(\bs)$. In the second step, we show that the expected total reward collected by the optimal policy is upper bounded by $(1+L)\cdot \sum_{\bs\in\mathcal{S}}p_1(\bs)\cdot\sum_{j\in[n]}\nu^1_j(\bs)$, which would imply a $1/(1+L)$ approximation ratio of our policy. The key point in finishing the second step is to use the bid price $\nu^t_j(\bs)$ described in \eqref{eqn:042202} to construct a feasible solution to LP \eqref{lp:ADP}. Then, with \Cref{lem:LPupperbound}, we prove our approximation ratio bound.

\subsection{Lower Bound the Total Reward of \Cref{alg:BMP}}

We now show that the total expected reward collected by \Cref{alg:BMP}, which is denoted by $\pi$, is lower bounded by $\sum_{\bs\in\mathcal{S}}p_1(\bs)\cdot\sum_{j\in[n]}\nu^1_j(\bs)$, with $\nu^t_j(\bs)$ defined in \eqref{eqn:042202} for for any $j\in[n]$, $t\in[T]$, and any $\bs\in\mathcal{S}$.

Note that for each period $t$ and each state $\bs_t$, we can view the bid price
\begin{equation}\label{eqn:042901}
\sum_{\bs'\in\mathcal{S}}p_t(\bs_t, \bs')\cdot\sum_{i\in\mathcal{A}_{j(\bs_t)}}\frac{1}{C_i}\cdot\sum_{j'\in\mathcal{B}_i}\nu^{t+1}_{j'}(\bs')
\end{equation}
as a threshold, and we only serve customer $t$ as long as its reward $r_{j(\bs_t)}$ exceeds this threshold and there are enough remaining capacities. Denote by $x_t^{\pi}(\bs_t)\in\{0,1\}$ the online decision made by our policy $\pi$ at period $t$ given state $\bs_t$. We decompose the total reward collected by policy $\pi$ into two parts based on the bid price \eqref{eqn:042901}. To be specific, we have
\begin{equation}\label{eqn:042902}
\begin{aligned}
\sum_{t\in[T]}\bI_{\{x^{\pi}_t(\bs_t)=1\}}\cdot r_{j(\bs_t)}=&\sum_{t\in[T]} \bI_{\{x^{\pi}_t(\bs_t)=1\}}\cdot \left[ r_{j(\bs_t)}-\sum_{\bs'\in\mathcal{S}}p_t(\bs_t, \bs')\cdot\sum_{i\in\mathcal{A}_{j(\bs_t)}}\frac{1}{C_i}\cdot\sum_{j'\in\mathcal{B}_i}\nu^{t+1}_{j'}(\bs') \right]\\
&+\sum_{t\in[T]} \bI_{\{x^{\pi}_t(\bs_t)=1\}}\cdot \left( \sum_{\bs'\in\mathcal{S}}p_t(\bs_t, \bs')\cdot\sum_{i\in\mathcal{A}_{j(\bs_t)}}\frac{1}{C_i}\cdot\sum_{j'\in\mathcal{B}_i}\nu^{t+1}_{j'}(\bs') \right).
\end{aligned}
\end{equation}
The first term in the right-hand side of \eqref{eqn:042902} can be further simplified with the decision rule \eqref{eqn:042904}, and the second term in the right-hand side of \eqref{eqn:042902} can be further simplified with the bid price update \eqref{eqn:042202}. Therefore, we can finally derive that
\[
\mathbb{E}\left[ \sum_{t\in[T]}\bI_{\{x^{\pi}_t(\bs_t)=1\}}\cdot r_{j(\bs_t)} \right]\geq \sum_{\bs\in\mathcal{S}}p_1(\bs)\cdot \sum_{j\in[n]}\nu^1_j(\bs).
\]
The above arguments are summarized in the following lemma, with the formal proof relegated to \Cref{sec:pf5}.
\begin{lemma}\label{lem:lowerbound}
The expected total reward collected by our \Cref{alg:BMP} is lower bounded by
\[
\sum_{\bs\in\mathcal{S}}p_1(\bs)\cdot \sum_{j\in[n]}\nu^1_j(\bs).
\]
\end{lemma}

\begin{remark}
The proof idea of \Cref{lem:lowerbound} is motivated by the analysis of the thresholding algorithm for prophet inequality (e.g. \cite{krengel1978semiamarts}), for NRM problem with reusable resources (e.g. \cite{baek2022bifurcating}), and for other stochastic online optimization problems (e.g. \cite{dutting2020prophet}), where we regard the bid price as a threshold for each type of customer based on the state, and we analyze the bid price part and the reward beyond the bid price part separately. An alternative idea to prove \Cref{lem:lowerbound} is based on the DP formulation, where we introduce a basis function $\psi_j(\cdot)$ for each type $j\in[n]$ and we use a linear combination of $\psi_j(\cdot)$ over $j\in[n]$ to lower bound the total revenue collected by \Cref{alg:BMP}. The bid price $\nu^t_j(\bs)$ now becomes the coefficient of the basis function $\psi_j(\cdot)$. Such an idea has been developed in \cite{ma2020approximation} for independent customer arrivals and we develop here for correlated customer arrivals, with further details referred to \Cref{sec:pf5}.
Note that the basis functions are only theoretically needed to prove a lower bound of the total reward collected by \Cref{alg:BMP}. The implementation of our \Cref{alg:BMP} does not require constructing any basis function. This is another difference (besides the bid price computing) between our \Cref{alg:BMP} and the approximate policy in \cite{ma2020approximation}.
\end{remark}

\subsection{Construct a Feasible Solution to LP \eqref{lp:ADP}}\label{sec:Construct}

We now construct a feasible solution to LP \eqref{lp:ADP} based on the bid price $\nu^t_j(\bs)$ described in \eqref{eqn:042202}. To be specific, we define
\begin{equation}\label{eqn:042301}
    \hat{\beta}^t_i(\bs)=\frac{1}{C_i}\cdot \sum_{j\in\mathcal{B}_i} \nu^t_j(\bs)
\end{equation}
for any $i\in[m]$, any $t\in[T]$ and any $\bs\in\mathcal{S}$. Also, starting from $\hat{\theta}^{T+1}(\bs)=0$ for any $\bs\in\mathcal{S}$, we iteratively define
\begin{equation}\label{eqn:042302}
    \hat{\theta}^t(\bs)=\sum_{j\in[n]}\nu^t_{j}(\bs)
\end{equation}
for any $\bs\in\mathcal{S}$, for $t=T, T-1, \dots, 1$. We now provide intuition on why $\{\hat{\beta}^t_i(\bs), \hat{\theta}^t(\bs), \forall i\in[m], \forall t\in[T], \forall \bs\in\mathcal{S}\}$ defined in \eqref{eqn:042301} and \eqref{eqn:042302} is feasible to LP \eqref{lp:ADP}. Note that from the definition in \eqref{eqn:042202}, we must have $\nu^t_j(\bs)\geq\sum_{\bs'\in\mathcal{S}}p_t(\bs, \bs')\cdot\nu^{t+1}_j(\bs')$ for any $j\in[n]$, $t\in[T]$ and $\bs\in\mathcal{S}$, which implies that
\[
\hat{\beta}^t_i(\bs)\geq\sum_{\bs'\in\mathcal{S}}p_t(\bs, \bs')\cdot\hat{\beta}^{t+1}_i(\bs').
\]
Therefore, in order for the constraint \eqref{const:ADP} to be satisfied, it is sufficient to show that
\[
 \hat{\theta}^t(\bs)-\sum_{\bs'\in\mathcal{S}}p_t(\bs, \bs')\cdot\hat{\theta}^{t+1}(\bs')\geq \left[ r_{j(\bs)}-\sum_{\bs'\in\mathcal{S}}p_t(\bs, \bs')\cdot\sum_{i\in[m]}a_{i,j(\bs)}\cdot \hat{\beta}^{t+1}_{i}(\bs') \right]^+.
\]
The above inequality can be directly verified by plugging in the definitions in \eqref{eqn:042301} and \eqref{eqn:042302}, and using the backward induction defined in \eqref{eqn:042202}. On the other hand, we can show that
\[
\sum_{\bs\in\mathcal{S}} p_1(\bs)\cdot \left( \hat{\theta}^1(\bs)+\sum_{i\in[m]}C_i\cdot\hat{\beta}^1_{i}(\bs) \right)\leq(1+L)\cdot \sum_{\bs\in\mathcal{S}} p_1(\bs)\cdot\sum_{j\in[n]}\nu^1_j(\bs)
\]
with parameter $L$ defined in \eqref{eqn:042203}. We summarize the above arguments in the following lemma, with the formal proof relegated to \Cref{sec:pf5}.
\begin{lemma}\label{lem:upperbound}
The set of solution $\{\hat{\beta}^t_i(\bs), \hat{\theta}^t(\bs), \forall i\in[m], \forall t\in[T], \forall \bs\in\mathcal{S}\}$ defined in \eqref{eqn:042301} and \eqref{eqn:042302} is feasible to LP \eqref{lp:ADP}. Moreover, it holds that
\[
\sum_{\bs\in\mathcal{S}} p_1(\bs)\cdot \left( \hat{\theta}^1(\bs)+\sum_{i\in[m]}C_i\cdot\hat{\beta}^1_{i}(\bs) \right)\leq(1+L)\cdot \sum_{\bs\in\mathcal{S}} p_1(\bs)\cdot\sum_{j\in[n]}\nu^1_j(\bs)
\]
with $L$ defined in \eqref{eqn:042203}.
\end{lemma}

\subsection{Proof of the Approximation Ratio Bound}\label{sec:proofCR}

We are now ready to prove the approximation ratio bound of \Cref{alg:BMP}. Our final result is formalized in the following theorem.

\begin{theorem}\label{thm:CR}
Let the parameter $L$ be defined in \eqref{eqn:042203}. Denote by $\ALG$ the expected total reward collected by \Cref{alg:BMP} and denote by $\OPT$ the expected total reward collected by the optimal policy. Then, it holds that
\[
\ALG\geq\frac{1}{1+L}\cdot\OPT.
\]
\end{theorem}
\begin{myproof}
From \Cref{lem:lowerbound}, we know that
\[
\ALG\geq \sum_{\bs\in\mathcal{S}}p_1(\bs)\cdot\sum_{j\in[n]}\nu^1_j(\bs)\]
with $\nu^t_j(\bs)$ defined in \eqref{eqn:042202} for for any $j\in[n]$, $t\in[T]$, and any $\bs\in\mathcal{S}$. Moreover, from \Cref{lem:LPupperbound} and \Cref{lem:upperbound}, we know that
\[
\OPT\leq \text{LP~}\eqref{lp:ADP}\leq \sum_{\bs\in\mathcal{S}} p_1(\bs)\cdot \left( \hat{\theta}^1(\bs)+\sum_{i\in[m]}C_i\cdot\hat{\beta}^1_{i}(\bs) \right)\leq(1+L)\cdot \sum_{\bs\in\mathcal{S}} p_1(\bs)\cdot\sum_{j\in[n]}\nu^1_j(\bs)
\]
with $\{\hat{\beta}^t_i(\bs), \hat{\theta}^t(\bs), \forall i\in[m], \forall t\in[T], \forall \bs\in\mathcal{S}\}$ defined in \eqref{eqn:042301} and \eqref{eqn:042302}. Our proof is thus completed.
\end{myproof}

Note that the approximation ratio bound established in \Cref{thm:CR} depends on $L$. Indeed, our problem can be reduced to the set packing problem studied in \cite{hazan2006complexity} by having \textit{deterministic} customer arrivals at each period. Then, it has been shown in Theorem 1 of \cite{hazan2006complexity} that even if the online policy has more power to make revocable decisions, which further reduces the problem into an offline problem, it is NP-hard to approximate the optimal policy with an approximation ratio bound better than $\Omega(\log L/L)$. Therefore, we know that it is inevitable to have the parameter $L$ showing up in our approximation ratio bound.

\section{Extention to the Choice-based Model}\label{sec:Choicemodel}

In this section, we extend our model and the performance guarantee to incorporate the choice model behavior of the customers. In the original formulation in \Cref{sec:problem}, each customer requires only one product and the decision maker only needs to decide whether or not to provide the requested product to the customer. We now consider a more general setting where the decision maker offers a subset $A_t\in \mathcal{F}$ of products at each period $t$, where $\mathcal{F}$ denotes the collection of all feasible assortments which contains the empty set $\emptyset$, and the customer would choose one product from $A_t$ according to its underlying choice model. Note that the customer can leave without purchase and we introduce a null product with $0$ reward and $0$ consumption of the resources to incorporate the leaving behavior of the customer. We require every assortment $A_t$ to contain the null product, which is indexed by product $0$. The choice probabilities now depend on the system state $\bs_t$. To be more concrete, we denote by $\phi_j(A_t, \bs_t)$ the probability that customer $t$ would choose product $j\in[n]$ to purchase, given the offered assortment $A_t$ and the system state $\bs_t$. Since the null product is always included in the assortment $A_t$, we have $\sum_{j\in A}\phi_j(A, \bs)=1$. We assume that the probabilities $\{ \phi_j(A, \bs), \forall j\in[n], \forall A\in\mathcal{F}, \forall \bs\in\mathcal{S} \}$ are given and the goal of the decision maker is to choose the assortment $A_t$ at each period $t\in[T]$ to maximize the total collected reward, subject to the capacity constraints of the resources. We adopt the standard substitutability assumption (e.g. \cite{golrezaei2014real}) over the assortments.

\begin{assumption}\label{assump:assort}
For any system state $\bs\in\mathcal{S}$, any assortment $A\in\mathcal{F}$, any product $j\in A$ and product $j'\notin A$, it holds that $\phi_j(A, \bs)\geq \phi_j(A\cup\{j'\}, \bs)$. Moreover, if $A\in\mathcal{F}$, then for any subset $B\subset A$, it holds $B\in\mathcal{F}$.
\end{assumption}

Denote by $V_{t}^*(\bc, \bs)$ the value to go function at period $t$, given the remaining capacity $\bc$ and the current state $\bs$. The backward induction can be given as follows:
\begin{equation}\label{eqn:Assortbackward}
\begin{aligned}
V^*_{t}(\bc, \bs)=&\max_{A\in\mathcal{F}(\bc)}\left\{ \sum_{j\in A}\phi_j(A, \bs)\cdot\left( r_{j}+\sum_{\bs'\in\mathcal{S}}p_t(\bs, \bs')\cdot V^*_{t+1}(\bc-\ba_{j}, \bs')  \right)
 \right\},
\end{aligned}
\end{equation}
where $\mathcal{F}(\bc)\subset\mathcal{F}$ denotes the collection of assortments that we have enough remaining capacities for every product contained in the assortment. Again, we adopt the linear approximation \eqref{eqn:approximate} to approximate the value of $V^*_{t}(\bc, \bs)$ and the upper bound LP \eqref{lp:ADP} can now be reformulated as follows.
\begin{subequations}\label{lp:AssortADP}
\begin{align}
\min ~~&\sum_{\bs\in\mathcal{S}} p_1(\bs)\cdot \left( \theta^1(\bs)+\sum_{i\in[m]}C_i\cdot\beta^1_{i}(\bs) \right)\label{Ob:AssortADP}\\
\mbox{s.t.}~~& \theta^t(\bs)-\sum_{\bs'\in\mathcal{S}}p_t(\bs, \bs')\cdot\theta^{t+1}(\bs')\geq \sum_{j\in A}\phi_j(A, \bs)\cdot \left[ r_{j}-\sum_{\bs'\in\mathcal{S}}p_t(\bs, \bs')\cdot\sum_{i\in[m]}a_{i,j}\cdot \beta^{t+1}_{i}(\bs') \right]^+  \nonumber\\
&~~~~~~~~+\sum_{i\in[m]}C_i\cdot\left[ \sum_{\bs'\in\mathcal{S}}p_t(\bs, \bs')\cdot \beta^{t+1}_{i}(\bs')-\beta_{i}^t(\bs) \right]^+, ~~~~~\forall t\in[T], \forall \bs\in\mathcal{S}, \forall A\in\mathcal{F}  \label{const:AssortADP}\\
&\theta^t(\bs)\geq0, \beta^t_{i}(\bs)\geq0, \forall i\in[m], ~~~~~\forall t\in[T], \forall \bs\in\mathcal{S}.
\end{align}
\end{subequations}
We show in the following lemma that the optimal objective value of LP \eqref{lp:AssortADP} is an upper bound of the DP value $\sum_{\bs\in\mathcal{S}}p_1(\bs)\cdot V^*_1(\bm{C}, \bs)$, with formal proof relegated to \Cref{sec:pf6}.
\begin{lemma}\label{lem:AssortUB}
It holds that LP \eqref{lp:AssortADP} $\geq \sum_{\bs\in\mathcal{S}}p_1(\bs)\cdot V^*_1(\bm{C}, \bs)$.
\end{lemma}

We now derive our policy. We still assign a bid price $\nu^t_j(\bs)$ for each product $j\in[n]$, each state $\bs\in\mathcal{S}$ and each period $t\in[T]$. The bid price $\nu^t_{j}(\bs)$ is computed in a backward induction from $T$ to $1$. To be specific, we set $\nu^{T+1}_{j}(\bs)=0$ for any $j\in[n]$ and any $\bs\in\mathcal{S}$. Then, for $t=T, T-1,\dots, 1$, we compute iteratively
\begin{equation}\label{eqn:052202}
    \nu^t_j(\bs)=\sum_{\bs'\in\mathcal{S}}p_t(\bs, \bs')\cdot\nu^{t+1}_j(\bs')+\sum_{j\in \hat{A}_t(\bs)}\phi_j(\hat{A}_t(\bs), \bs) \cdot\left[r_{j}-\sum_{\bs'\in\mathcal{S}}p_t(\bs, \bs')\cdot\sum_{i\in\mathcal{A}_j}\frac{1}{C_i}\cdot\sum_{j'\in\mathcal{B}_i}\nu^{t+1}_{j'}(\bs') \right]^+
\end{equation}
for every $j\in[n]$ and every $\bs\in\mathcal{S}$, where the set $\hat{A}_t(\bs)$ is defined as follows
\begin{equation}\label{eqn:050701}
    \hat{A}_t(\bs)=\text{argmax}_{A\in\mathcal{F}}\left\{\sum_{j\in A}\phi_j(A, \bs) \cdot\left[r_{j}-\sum_{\bs'\in\mathcal{S}}p_t(\bs, \bs')\cdot\sum_{i\in\mathcal{A}_j}\frac{1}{C_i}\cdot\sum_{j'\in\mathcal{B}_i}\nu^{t+1}_{j'}(\bs') \right]^+\right\}.
\end{equation}

Our policy is formally described in \Cref{alg:assortBMP}.
\begin{algorithm}[ht!]
\caption{Assortment Bid Price Policy}
\label{alg:assortBMP}
\begin{algorithmic}[1]
\State Compute the bid prices $\nu^t_j(\bs)$ for any $j\in[n]$, any $t\in[T]$ and any $\bs\in\mathcal{S}$, following the backward induction equation described in \eqref{eqn:052202}.
\For{t=1,\dots,T}
\State observe the state $\bs_t$ and the remaining capacities $\bc_t$.
\State compute the assortment $\hat{A}_t(\bs)$ as defined in \eqref{eqn:050701}.
\State offer the assortment
\begin{equation}\label{eqn:050702}
   A_t(\bs_t)=\{ j\in \hat{A}_t(\bs_t): \bc_t\geq \ba_j  \}
\end{equation}
to customer $t$.
\EndFor
\end{algorithmic}
\end{algorithm}
We have the following approximation ratio bound regarding the policy in \Cref{alg:assortBMP}, with the formal proof relegated to \Cref{sec:pf6}.

\begin{theorem}\label{thm:assortCR}
Let the parameter $L$ be defined in \eqref{eqn:042203}. Denote by $\ALG$ the expected total reward collected by \Cref{alg:assortBMP} and denote by $\OPT$ the expected total reward collected by the optimal policy. Then, it holds that
\[
\ALG\geq\frac{1}{1+L}\cdot\OPT.
\]
\end{theorem}

\section{Numerical Studies}
In this section, we conduct numerical studies to present the empirical performance of our algorithms. Beyond \Cref{alg:BMP} where the bid prices are computed iteratively via the backward induction \eqref{eqn:042202}, we also develop another approximate DP heuristics where the bid prices are computed from solving the LP \eqref{lp:ADP}. In what follows, we first develop the approximate DP heuristics, and we then present the numerical results.

\subsection{An Approximate DP Heuristics}
Note that a well-studied approximate DP heuristics can be derived via the following procedure (e.g. \cite{adelman2007dynamic}). We first solve the LP formulation of the approximate DP, which is LP \eqref{lp:ADP} in our setting, to obtain one set of the optimal parameters $\{\theta^{t*}(\bs), 
 \beta^{t*}_i(\bs), \forall t\in[T], \forall \bs\in\mathcal{S}, \forall i\in[m]\}$ to be used in the linear approximation \eqref{eqn:approximate}. We then embed the linear approximation \eqref{eqn:approximate} with the set of parameters $\{\theta^{t*}(\bs), 
 \beta^{t*}_i(\bs), \forall t\in[T], \forall \bs\in\mathcal{S}, \forall i\in[m]\}$ into the DP decision rule \eqref{eqn:042601}, and we obtain the following decision rule based on our linear approximation: given the state $\bs_t$, we serve customer $t$ as long as $\bc_t\geq\ba_{j(\bs_t)}$ and
\begin{equation}\label{eqn:100501}
r_{j(\bs_t)}\geq \sum_{\bs'\in\mathcal{S}}p_t(\bs_t, \bs')\cdot \sum_{i\in[m]} a_{i, j(\bs_t)}\cdot \beta_i^{(t+1)*}(\bs').
\end{equation}
The formal algorithm is presented in \Cref{alg:linearADP}. Note that the decision rule \eqref{eqn:100501} can actually be described in a more general way, where we can introduce a set of basis functions $\{\phi_k(\bc)\}_{k=1}^K$ and approximate the DP value by 
\begin{equation}\label{eqn:100502}
V_{t}(\bc, \bs)=\theta^t(\bs)+\sum_{k\in[K]}\phi_k(\bc)\cdot \beta_{k}^t(\bs),~~\forall \bc.
\end{equation}
After the set of parameters $\left\{\theta^t(\bs), \beta^t_{k}(\bs), \forall k\in[K], \forall \bs\in\mathcal{S}, \forall t\in[T] \right\}$ is determined, we may adopt the following decision rule: 
given the state $\bs_t$, we serve customer $t$ as long as $\bc_t\geq\ba_{j(\bs_t)}$ and
\begin{equation}\label{eqn:100503}
r_{j(\bs_t)}\geq \sum_{\bs'\in\mathcal{S}}p_t(\bs_t, \bs')\cdot \sum_{k\in[K]} \beta^t_k(\bs')\cdot\left(\phi_k(\bc_t)-\phi_k(\bc_t-\ba_{j(\bs_t)})\right).
\end{equation}
When $K=m$ and for each $k\in[m]$, $\phi_k(\bc)=c_k$, the decision rule \eqref{eqn:100503} recovers the decision rule \eqref{eqn:100501}. We may also consider using a concave function as the basis function (e.g. \cite{farias2007approximate}) or using a neural network as the basis function (the weight parameters $\beta_k^t$ will be absorbed into the design of the neural network), for example, in \cite{bertsekas1995neuro}. The numerical results we present are for the linear approximation with the decision rule \eqref{eqn:100501}. We leave the development of the approximate DP heuristics with more involved basis functions under correlated customer arrivals for future research.

\begin{algorithm}[ht!]
\caption{Approximate DP Heuristics}
\label{alg:linearADP}
\begin{algorithmic}[1]
\State Solve the LP formulation of the approximate DP, which is LP \eqref{lp:ADP}, to obtain one set of the optimal parameters $\{\theta^{t*}(\bs), 
 \beta^{t*}_i(\bs), \forall t\in[T], \forall \bs\in\mathcal{S}, \forall i\in[m]\}$
\For{t=1,\dots,T}
\State observe the state $\bs_t$ and the remaining capacities $\bc_t$.
\If {there exists a resource $i\in[m]$ such that $c_{t,i}< a_{i, j(\bs)}$} {reject customer $t$}
\Else {~serve customer $t$ if and only if \begin{equation}\label{eqn:100504} 
r_{j(\bs_t)}\geq \sum_{\bs'\in\mathcal{S}}p_t(\bs_t, \bs')\cdot \sum_{i\in[m]} a_{i, j(\bs_t)}\cdot \beta_i^{(t+1)*}(\bs').
\end{equation}}
\EndIf
\EndFor
\end{algorithmic}
\end{algorithm}

\subsection{Experiment Setup and Results}
In this section, we describe the setup for our numerical experiments and we present our numerical results over the performance of \Cref{alg:BMP} and \Cref{alg:linearADP}, compared to the objective value of LP \eqref{lp:ADP}, which serves as an upper bound of the total reward collected by the optimal policy.

We consider an airline revenue management example. There are 5 locations in the network, including one hub and four spokes. There is a flight leg from each spoke to the hub, and from the hub to each spoke. As a result, in total, there are 8 flight legs. The customer can be required to travel from each location to any other location, which means there are 20 origin-destination pairs. However, since there is no flight leg directly connecting the two spokes, the customer traveling between spokes needs to transfer through the hub and thus requires two flight legs. For each origin-destination pair, there is a high-fare customer class and a low-fare customer class. Therefore, there are 40 customer classes in total. The experimental setup described above is standard in the literature and has been widely adopted by previous papers (e.g. \cite{topaloglu2009using, hu2013revenue, brown2014information, vossen2015dynamic, vossen2015reductions, kunnumkal2016piecewise, ma2020approximation, li2023revenue}).

For each origin-destination pair, we first randomly generate a reward for the low-fare customer class following a uniform distribution over $[0,1]$. We then set the reward for the high-fare customer class to be two times the reward for the corresponding low-fare customer class. We now describe how the parameters regarding the customer arrival process of the experiments are selected. We first denote a random variable $D\sim\mathcal{N}(\mu, \sigma)$, following a normal distribution with mean $\mu$ and variance $\sigma$. We set the total number of periods in the horizon $T$ to be the smallest integer such that $P(D\leq T)\geq 0.9$. Then, for each $t=1,\dots, T$, we define $\rho_t=P(D\geq t+1|D\geq t)$ to be the probability that customer $t+1$ arrives conditioning on customer $t$ arrives. We then try two different ways of generating the probabilities for each customer to be realized as one type.
\begin{itemize}
    \item Setting A: For each $t=1,\dots, T$, conditioning on customer $t$ arrives, we let customer $t$ require one origin-destination pair, indexed by $i$, with probability $\lambda_{i,t}=a_{i,t}/\sum_{i'} a_{i',t}$, where each $a_{i',t}$ is uniformly drawn from $[0,1]$. Then, conditioning on customer $t$ requires one origin-destination pair, it belongs to the high-fare customer class with probability $\frac{1}{2}+\frac{t}{2T}$ and belongs to the low-fare customer class with probability $\frac{1}{2}-\frac{t}{2T}$. Such a setting reflects the fact that the customers arriving near the end of the horizon are more willing to pay a higher price.
    \item Setting B: For each $t=1,\dots, T$, conditioning on customer $t-1$ arrives requiring an origin-destination pair $\hat{i}$ and customer $t$ arrives, we let customer $t$ require one origin-destination pair, indexed by $i$, with probability $\lambda_{\hat{i},i,t}=a_{\hat{i}, i,t}/\sum_{i'} a_{\hat{i}, i',t}$, where each $a_{\hat{i},i',t}$ is uniformly drawn from $[0,1]$. Then, conditioning on customer $t$ requires one origin-destination pair, customer $t$ belongs to the high-fare customer class with probability $\frac{1}{2}+\frac{t}{2T}$ and belongs to the low-fare customer class with probability $\frac{1}{2}-\frac{t}{2T}$.
\end{itemize}
Note that setting B described above is more involved than setting A in that the probabilities of the type realization of customer $t$ also depend on the type realization of customer $t-1$ in setting B. In contrast, in setting A, only the arrival of customer $t$ depends on whether customer $t-1$ arrives and the type realization of customer $t$ is independent of the type of customer $t-1$.

The numerical results for setting A are presented in \Cref{tab:settingA}. We implement \Cref{alg:BMP}, \Cref{alg:linearADP} and compute the value of LP \eqref{lp:ADP} for different choices of $(\mu, \sigma)$. In order to compute the expected cost for each algorithm, we repeat running the algorithm for $K=1000$ times and take the average. As we can see, the performances of \Cref{alg:BMP} and \Cref{alg:linearADP} are both quite good compared to the optimal policy. In most cases, the gap between the total reward collected by \Cref{alg:BMP} and the average total reward collected by \Cref{alg:linearADP}, relative to the value of the upper bound LP \eqref{lp:ADP}, are within $10\%$. On average, the gap between the total reward collected by \Cref{alg:BMP} relative to the value of LP \eqref{lp:ADP} is $6.60\%$. The average gap between the total reward collected by \Cref{alg:linearADP} relative to the value of LP \eqref{lp:ADP} is $8.08\%$. Similar results hold for setting B, as illustrated in \Cref{tab:settingB}. On average, the gap between the total reward collected by \Cref{alg:BMP} relative to the value of LP \eqref{lp:ADP} is $6.60\%$. The gap between the total reward collected by \Cref{alg:linearADP} relative to the value of LP \eqref{lp:ADP} is $6.62\%$. These experiments show that both \Cref{alg:BMP} and \Cref{alg:linearADP} perform reasonably well compared to the optimal policy.

\begin{table}[h]
    \centering
    \begin{tabular}{|c|c|c|c|c|c|c|}
    \hline
       $(\mu, \sigma)$  & value of $T$ & Upper Bound & BBP & ADP heuristics & BBP gap ratio & ADP heuristics gap ratio  \\
\hline
(30, 15) & 56 & 17.94 & 15.87 & 16.25 & $11.52\%$ & $9.42\%$ \\
\hline
(40, 15) & 66 & 37.29 & 34.46 & 34.51 & $7.59\%$ & $7.45\%$ \\
\hline
(50, 15) & 76 & 55.74 & 52.29 & 50.95 & $6.19\%$ & $8.60\%$ \\
\hline
(60, 15) & 86 & 52.57 & 50.33 & 50.42 & $4.26\%$ & $4.10\%$ \\
\hline
(40, 10) & 53 & 34.51 & 32.25 & 32.32 & $6.52\%$ & $6.33\%$ \\
\hline
(40, 15) & 66 & 28.55 & 27.15 & 24.73 & $4.90\%$ & $13.40\%$ \\
\hline
(40, 25) & 79 & 37.65 & 35.46 & 35.01 & $5.80\%$ & $7.01\%$ \\
\hline
(40, 30) & 92 & 29.65 & 27.85 & 27.18 & $6.07\%$ & $8.33\%$ \\
\hline
    \end{tabular}
    \caption{Numerical results for setting A. In the table, ``Upper Bound'' stands for the value of LP \eqref{lp:ADP}. ``BBP'' stands for the expected total reward collected by \Cref{alg:BMP}. ``ADP heuristics'' stands for the expected total reward collected by \Cref{alg:linearADP}. ``BBP gap ratio'' denotes the value of $(\text{Upper~Bound}-\text{BBP})/\text{Upper~Bound}$. ``ADP heuristics gap ratio'' denotes the value of $(\text{Upper~Bound}-\text{ADP heuristics})/\text{Upper~Bound}$.}
    \label{tab:settingA}
\end{table}

\begin{table}[h]
    \centering
    \begin{tabular}{|c|c|c|c|c|c|c|}
    \hline
       $(\mu, \sigma)$  & value of $T$ & Upper Bound & BBP & ADP heuristics & BBP ratio & ADP heuristics ratio  \\
\hline
(30, 15) & 56 & 26.32 & 23.34 & 23.32 & $11.34\%$ & $11.41\%$ \\
\hline
(40, 15) & 66 & 26.46 & 25.33 & 24.53 & $4.26\%$ & $7.27\%$ \\
\hline
(50, 15) & 76 & 40.80 & 37.35 & 38.15 & $8.44\%$ & $6.48\%$ \\
\hline
(60, 15 ) & 86 & 47.17 & 45.92 & 45.25 & $2.67\%$ & $4.07\%$ \\
\hline
(40, 10) & 53 & 32.07 & 29.66 & 29.58 & $7.49\%$ & $7.76\%$ \\
\hline
(40, 15) & 66 & 30.38 & 29.07 & 29.14 & $4.29\%$ & $4.06\%$ \\
\hline
(40, 25) & 79 & 38.55 & 35.86 & 35.57 & $6.99\%$ & $7.74\%$ \\
\hline
(40, 30) & 92 & 33.12 & 30.70 & 31.75 & $7.31\%$ & $4.16\%$ \\
\hline
    \end{tabular}
    \caption{Numerical results for setting B. In the table, ``Upper Bound'' stands for the value of LP \eqref{lp:ADP}. ``BBP'' stands for the expected total reward collected by \Cref{alg:BMP}. ``ADP heuristics'' stands for the expected total reward collected by \Cref{alg:linearADP}. ``BBP ratio'' denotes the value of $(\text{Upper~Bound}-\text{BBP})/\text{Upper~Bound}$. ``ADP heuristics ratio'' denotes the value of $(\text{Upper~Bound}-\text{ADP heuristics})/\text{Upper~Bound}$.}
    \label{tab:settingB}
\end{table}

\section{Concluding Remarks}

We consider the NRM problem with correlated customer arrivals. Our contributions are threefold. First, we propose a new model that assumes the existence of a system state, which determines customer arrivals for the current period. This system state evolves over time according to a time-inhomogeneous Markov chain. Our model can be used to represent correlation in various settings. Second, we develop a new LP approximation of the optimal policy. Our approximation is motivated from the approximate DP literature and serves as an asymptotically optimal upper bound over the expected total reward collected by the optimal policy. Third, we develop a new bid price policy and show that our policy enjoys an approximation ratio bound of $1/(1+L)$. Then, we extend all our results to the assortment setting where the decision maker could offer an assortment to the customer at each period and the customer would choose one product to purchase according to its underlying choice model. This extension is important because it captures the general scenario where customers have different preferences and can choose from a variety of products. Finally, we propose another approximate DP heuristics and study both our algorithms empirically. Through the numerical experiments, we show that the performances of both our algorithms are reasonably well compared to the optimal policy. There are also multiple directions to further extend our results. For example, one may consider a reusable resource setting where each unit of resource can be returned after a certain period. One may also consider an overbooking setting where the customers may reserve some resources but eventually do not consume the resources. We leave these interesting topics for future research.

\ACKNOWLEDGMENT{We thank Stefanus Jasin and Billy Jin for helpful discussions of the project. We also would like to thank Rajan Udwani and Huseyin Topaloglu for their helpful feedbacks and comments on the paper.
}

\bibliographystyle{abbrvnat}
\bibliography{bibliography}

\begin{thebibliography}{61}
\providecommand{\natexlab}[1]{#1}
\providecommand{\url}[1]{\texttt{#1}}
\expandafter\ifx\csname urlstyle\endcsname\relax
  \providecommand{\doi}[1]{doi: #1}\else
  \providecommand{\doi}{doi: \begingroup \urlstyle{rm}\Url}\fi

\bibitem[Adelman(2007)]{adelman2007dynamic}
D.~Adelman.
\newblock Dynamic bid prices in revenue management.
\newblock \emph{Operations Research}, 55\penalty0 (4):\penalty0 647--661, 2007.

\bibitem[Altman(1999)]{altman1999constrained}
E.~Altman.
\newblock \emph{Constrained Markov decision processes}, volume~7.
\newblock CRC press, 1999.

\bibitem[Aouad and Ma(2022)]{aouad2022nonparametric}
A.~Aouad and W.~Ma.
\newblock A nonparametric framework for online stochastic matching with
  correlated arrivals.
\newblock \emph{arXiv preprint arXiv:2208.02229}, 2022.

\bibitem[Aviv and Pazgal(2005)]{aviv2005partially}
Y.~Aviv and A.~Pazgal.
\newblock A partially observed markov decision process for dynamic pricing.
\newblock \emph{Management science}, 51\penalty0 (9):\penalty0 1400--1416,
  2005.

\bibitem[Baek and Ma(2022)]{baek2022bifurcating}
J.~Baek and W.~Ma.
\newblock Bifurcating constraints to improve approximation ratios for network
  revenue management with reusable resources.
\newblock \emph{Operations Research}, 2022.

\bibitem[Bai et~al.(2022)Bai, El~Housni, Jin, Rusmevichientong, Topaloglu, and
  Williamson]{bai2022fluid}
Y.~Bai, O.~El~Housni, B.~Jin, P.~Rusmevichientong, H.~Topaloglu, and D.~P.
  Williamson.
\newblock Fluid approximations for revenue management under high-variance
  demand.
\newblock \emph{Management science (forthcoming)}, 2022.

\bibitem[Balseiro et~al.(2021)Balseiro, Besbes, and
  Pizarro]{balseiro2021survey}
S.~Balseiro, O.~Besbes, and D.~Pizarro.
\newblock Survey of dynamic resource constrained reward collection problems:
  Unified model and analysis.
\newblock \emph{Operations Research (forthcoming)}, 2021.

\bibitem[Bertsekas and Tsitsiklis(1995)]{bertsekas1995neuro}
D.~P. Bertsekas and J.~N. Tsitsiklis.
\newblock Neuro-dynamic programming: an overview.
\newblock In \emph{Proceedings of 1995 34th IEEE conference on decision and
  control}, volume~1, pages 560--564. IEEE, 1995.

\bibitem[Besbes et~al.(2022)Besbes, Kanoria, and
  Kumar]{besbes2022multisecretary}
O.~Besbes, Y.~Kanoria, and A.~Kumar.
\newblock The multisecretary problem with many types.
\newblock \emph{arXiv preprint arXiv:2205.09078}, 2022.

\bibitem[Bray(2022)]{bray2022logarithmic}
R.~Bray.
\newblock Logarithmic regret in multisecretary and online linear programming
  problems with continuous valuations.
\newblock \emph{arXiv preprint arXiv:1912.08917}, 2022.

\bibitem[Brown and Smith(2014)]{brown2014information}
D.~B. Brown and J.~E. Smith.
\newblock Information relaxations, duality, and convex stochastic dynamic
  programs.
\newblock \emph{Operations Research}, 62\penalty0 (6):\penalty0 1394--1415,
  2014.

\bibitem[Bumpensanti and Wang(2020)]{bumpensanti2020re}
P.~Bumpensanti and H.~Wang.
\newblock A re-solving heuristic with uniformly bounded loss for network
  revenue management.
\newblock \emph{Management Science}, 66\penalty0 (7):\penalty0 2993--3009,
  2020.

\bibitem[Chen and Song(2001)]{chen2001optimal}
F.~Chen and J.-S. Song.
\newblock Optimal policies for multiechelon inventory problems with
  markov-modulated demand.
\newblock \emph{Operations Research}, 49\penalty0 (2):\penalty0 226--234, 2001.

\bibitem[Chen et~al.(2019)Chen, Wen, and Xie]{chen2019dynamic}
Y.~Chen, Z.~Wen, and Y.~Xie.
\newblock Dynamic pricing in an evolving and unknown marketplace.
\newblock \emph{Available at SSRN 3382957}, 2019.

\bibitem[De~Farias and Van~Roy(2003)]{de2003linear}
D.~P. De~Farias and B.~Van~Roy.
\newblock The linear programming approach to approximate dynamic programming.
\newblock \emph{Operations research}, 51\penalty0 (6):\penalty0 850--865, 2003.

\bibitem[den Boer et~al.(2018)den Boer, Perry, and Zwart]{den2018dynamic}
A.~den Boer, O.~Perry, and B.~Zwart.
\newblock Dynamic pricing policies for an inventory model with random windows
  of opportunities.
\newblock \emph{Naval Research Logistics (NRL)}, 65\penalty0 (8):\penalty0
  660--675, 2018.

\bibitem[Dutting et~al.(2020)Dutting, Feldman, Kesselheim, and
  Lucier]{dutting2020prophet}
P.~Dutting, M.~Feldman, T.~Kesselheim, and B.~Lucier.
\newblock Prophet inequalities made easy: Stochastic optimization by pricing
  nonstochastic inputs.
\newblock \emph{SIAM Journal on Computing}, 49\penalty0 (3):\penalty0 540--582,
  2020.

\bibitem[Efroni et~al.(2020)Efroni, Mannor, and Pirotta]{efroni2020exploration}
Y.~Efroni, S.~Mannor, and M.~Pirotta.
\newblock Exploration-exploitation in constrained mdps.
\newblock \emph{arXiv preprint arXiv:2003.02189}, 2020.

\bibitem[Farias and Van~Roy(2007)]{farias2007approximate}
V.~F. Farias and B.~Van~Roy.
\newblock An approximate dynamic programming approach to network revenue
  management.
\newblock \emph{Preprint}, 2007.

\bibitem[Feng and Gallego(2000)]{feng2000perishable}
Y.~Feng and G.~Gallego.
\newblock Perishable asset revenue management with markovian time dependent
  demand intensities.
\newblock \emph{Management science}, 46\penalty0 (7):\penalty0 941--956, 2000.

\bibitem[Freund and Banerjee(2019)]{freund2019good}
D.~Freund and S.~Banerjee.
\newblock Good prophets know when the end is near.
\newblock \emph{Available at SSRN 3479189}, 2019.

\bibitem[Freund and Zhao(2022)]{freund2021overbooking}
D.~Freund and J.~Zhao.
\newblock Overbooking with bounded loss.
\newblock In \emph{Mathematics of Operations Research, forthcoming}, 2022.

\bibitem[Gallego and Van~Ryzin(1994)]{gallego1994optimal}
G.~Gallego and G.~Van~Ryzin.
\newblock Optimal dynamic pricing of inventories with stochastic demand over
  finite horizons.
\newblock \emph{Management science}, 40\penalty0 (8):\penalty0 999--1020, 1994.

\bibitem[Germano et~al.(2023)Germano, Stradi, Genalti, Castiglioni, Marchesi,
  and Gatti]{germano2023best}
J.~Germano, F.~E. Stradi, G.~Genalti, M.~Castiglioni, A.~Marchesi, and
  N.~Gatti.
\newblock A best-of-both-worlds algorithm for constrained mdps with long-term
  constraints.
\newblock \emph{arXiv preprint arXiv:2304.14326}, 2023.

\bibitem[Golrezaei et~al.(2014)Golrezaei, Nazerzadeh, and
  Rusmevichientong]{golrezaei2014real}
N.~Golrezaei, H.~Nazerzadeh, and P.~Rusmevichientong.
\newblock Real-time optimization of personalized assortments.
\newblock \emph{Management Science}, 60\penalty0 (6):\penalty0 1532--1551,
  2014.

\bibitem[Hazan et~al.(2006)Hazan, Safra, and Schwartz]{hazan2006complexity}
E.~Hazan, S.~Safra, and O.~Schwartz.
\newblock On the complexity of approximating k-set packing.
\newblock \emph{computational complexity}, 15\penalty0 (1):\penalty0 20--39,
  2006.

\bibitem[Hu et~al.(2013)Hu, Caldentey, and Vulcano]{hu2013revenue}
X.~Hu, R.~Caldentey, and G.~Vulcano.
\newblock Revenue sharing in airline alliances.
\newblock \emph{Management Science}, 59\penalty0 (5):\penalty0 1177--1195,
  2013.

\bibitem[Jasin and Kumar(2012)]{jasin2012re}
S.~Jasin and S.~Kumar.
\newblock A re-solving heuristic with bounded revenue loss for network revenue
  management with customer choice.
\newblock \emph{Mathematics of Operations Research}, 37\penalty0 (2):\penalty0
  313--345, 2012.

\bibitem[Jasin and Kumar(2013)]{jasin2013analysis}
S.~Jasin and S.~Kumar.
\newblock Analysis of deterministic lp-based booking limit and bid price
  controls for revenue management.
\newblock \emph{Operations Research}, 61\penalty0 (6):\penalty0 1312--1320,
  2013.

\bibitem[Jia et~al.(2023)Jia, Li, Liu, Liu, Zhou, Gravin, and
  Tang]{jia2023online}
J.~Jia, H.~Li, K.~Liu, Z.~Liu, J.~Zhou, N.~Gravin, and Z.~G. Tang.
\newblock Online resource allocation in markov chains.
\newblock In \emph{Proceedings of the ACM Web Conference 2023}, pages
  3498--3507, 2023.

\bibitem[Jiang et~al.(2020)Jiang, Li, and Zhang]{jiang2020online}
J.~Jiang, X.~Li, and J.~Zhang.
\newblock Online stochastic optimization with wasserstein based
  non-stationarity.
\newblock \emph{arXiv preprint arXiv:2012.06961}, 2020.

\bibitem[Jiang et~al.(2022)Jiang, Ma, and Zhang]{jiang2022degeneracy}
J.~Jiang, W.~Ma, and J.~Zhang.
\newblock Degeneracy is ok: Logarithmic regret for network revenue management
  with indiscrete distributions.
\newblock \emph{arXiv preprint arXiv:2210.07996}, 2022.

\bibitem[Keskin and Li(2022)]{keskin2022selling}
N.~B. Keskin and M.~Li.
\newblock Selling quality-differentiated products in a markovian market with
  unknown transition probabilities.
\newblock \emph{Available at SSRN 3526568}, 2022.

\bibitem[Krengel and Sucheston(1978)]{krengel1978semiamarts}
U.~Krengel and L.~Sucheston.
\newblock On semiamarts, amarts, and processes with finite value.
\newblock \emph{Probability on Banach spaces}, 4:\penalty0 197--266, 1978.

\bibitem[Kunnumkal and Talluri(2016)]{kunnumkal2016piecewise}
S.~Kunnumkal and K.~Talluri.
\newblock On a piecewise-linear approximation for network revenue management.
\newblock \emph{Mathematics of Operations Research}, 41\penalty0 (1):\penalty0
  72--91, 2016.

\bibitem[Li et~al.(2023)Li, Rusmevichientong, and Topaloglu]{li2023revenue}
W.~Li, P.~Rusmevichientong, and H.~Topaloglu.
\newblock Revenue management with calendar-aware and dependent demands:
  Asymptotically tight fluid approximations.
\newblock \emph{Available at SSRN 4543277}, 2023.

\bibitem[Ma et~al.(2020)Ma, Rusmevichientong, Sumida, and
  Topaloglu]{ma2020approximation}
Y.~Ma, P.~Rusmevichientong, M.~Sumida, and H.~Topaloglu.
\newblock An approximation algorithm for network revenue management under
  nonstationary arrivals.
\newblock \emph{Operations Research}, 68\penalty0 (3):\penalty0 834--855, 2020.

\bibitem[Meissner and Strauss(2012)]{meissner2012network}
J.~Meissner and A.~Strauss.
\newblock Network revenue management with inventory-sensitive bid prices and
  customer choice.
\newblock \emph{European Journal of Operational Research}, 216\penalty0
  (2):\penalty0 459--468, 2012.

\bibitem[Powell(2007)]{powell2007approximate}
W.~B. Powell.
\newblock \emph{Approximate Dynamic Programming: Solving the curses of
  dimensionality}, volume 703.
\newblock John Wiley \& Sons, 2007.

\bibitem[Qiu et~al.(2020)Qiu, Wei, Yang, Ye, and Wang]{qiu2020upper}
S.~Qiu, X.~Wei, Z.~Yang, J.~Ye, and Z.~Wang.
\newblock Upper confidence primal-dual reinforcement learning for cmdp with
  adversarial loss.
\newblock \emph{Advances in Neural Information Processing Systems},
  33:\penalty0 15277--15287, 2020.

\bibitem[Reiman and Wang(2008)]{reiman2008asymptotically}
M.~I. Reiman and Q.~Wang.
\newblock An asymptotically optimal policy for a quantity-based network revenue
  management problem.
\newblock \emph{Mathematics of Operations Research}, 33\penalty0 (2):\penalty0
  257--282, 2008.

\bibitem[Rustichini and Wolinsky(1995)]{rustichini1995learning}
A.~Rustichini and A.~Wolinsky.
\newblock Learning about variable demand in the long run.
\newblock \emph{Journal of economic Dynamics and Control}, 19\penalty0
  (5-7):\penalty0 1283--1292, 1995.

\bibitem[Schweitzer and Seidmann(1985)]{schweitzer1985generalized}
P.~J. Schweitzer and A.~Seidmann.
\newblock Generalized polynomial approximations in markovian decision
  processes.
\newblock \emph{Journal of mathematical analysis and applications},
  110\penalty0 (2):\penalty0 568--582, 1985.

\bibitem[Sethi and Cheng(1997)]{sethi1997optimality}
S.~P. Sethi and F.~Cheng.
\newblock Optimality of (s, s) policies in inventory models with markovian
  demand.
\newblock \emph{Operations Research}, 45\penalty0 (6):\penalty0 931--939, 1997.

\bibitem[Simchi-Levi et~al.(2004)Simchi-Levi, Wu, and Shen]{simchi2004handbook}
D.~Simchi-Levi, S.~D. Wu, and Z.-J.~M. Shen.
\newblock \emph{Handbook of quantitative supply chain analysis: modeling in the
  e-business era}, volume~74.
\newblock Springer Science \& Business Media, 2004.

\bibitem[Simchi-Levi et~al.(2022)Simchi-Levi, Zheng, and Zhu]{simchi2022online}
D.~Simchi-Levi, Z.~Zheng, and F.~Zhu.
\newblock Online matching with reusable network resources and decaying rewards:
  A unified framework.
\newblock \emph{Available at SSRN 3981123}, 2022.

\bibitem[Song and Zipkin(1993)]{song1993inventory}
J.-S. Song and P.~Zipkin.
\newblock Inventory control in a fluctuating demand environment.
\newblock \emph{Operations Research}, 41\penalty0 (2):\penalty0 351--370, 1993.

\bibitem[Talluri and Van~Ryzin(1998)]{talluri1998analysis}
K.~Talluri and G.~Van~Ryzin.
\newblock An analysis of bid-price controls for network revenue management.
\newblock \emph{Management science}, 44\penalty0 (11-part-1):\penalty0
  1577--1593, 1998.

\bibitem[Tong and Topaloglu(2014)]{tong2014approximate}
C.~Tong and H.~Topaloglu.
\newblock On the approximate linear programming approach for network revenue
  management problems.
\newblock \emph{INFORMS Journal on Computing}, 26\penalty0 (1):\penalty0
  121--134, 2014.

\bibitem[Topaloglu(2009)]{topaloglu2009using}
H.~Topaloglu.
\newblock Using lagrangian relaxation to compute capacity-dependent bid prices
  in network revenue management.
\newblock \emph{Operations Research}, 57\penalty0 (3):\penalty0 637--649, 2009.

\bibitem[Truong and Wang(2019)]{truong2019prophet}
V.-A. Truong and X.~Wang.
\newblock Prophet inequality with correlated arrival probabilities, with
  application to two sided matchings.
\newblock \emph{arXiv preprint arXiv:1901.02552}, 2019.

\bibitem[Vera and Banerjee(2021)]{vera2021bayesian}
A.~Vera and S.~Banerjee.
\newblock The bayesian prophet: A low-regret framework for online decision
  making.
\newblock \emph{Management Science}, 67\penalty0 (3):\penalty0 1368--1391,
  2021.

\bibitem[Vera et~al.(2021)Vera, Banerjee, and Gurvich]{vera2021online}
A.~Vera, S.~Banerjee, and I.~Gurvich.
\newblock Online allocation and pricing: Constant regret via bellman
  inequalities.
\newblock \emph{Operations Research}, 69\penalty0 (3):\penalty0 821--840, 2021.

\bibitem[Vossen and Zhang(2015{\natexlab{a}})]{vossen2015dynamic}
T.~W. Vossen and D.~Zhang.
\newblock A dynamic disaggregation approach to approximate linear programs for
  network revenue management.
\newblock \emph{Production and Operations Management}, 24\penalty0
  (3):\penalty0 469--487, 2015{\natexlab{a}}.

\bibitem[Vossen and Zhang(2015{\natexlab{b}})]{vossen2015reductions}
T.~W. Vossen and D.~Zhang.
\newblock Reductions of approximate linear programs for network revenue
  management.
\newblock \emph{Operations Research}, 63\penalty0 (6):\penalty0 1352--1371,
  2015{\natexlab{b}}.

\bibitem[Wang and Wang(2022)]{wang2022constant}
Y.~Wang and H.~Wang.
\newblock Constant regret resolving heuristics for price-based revenue
  management.
\newblock \emph{Operations Research}, 2022.

\bibitem[Wei et~al.(2018)Wei, Yu, and Neely]{wei2018online}
X.~Wei, H.~Yu, and M.~J. Neely.
\newblock Online learning in weakly coupled markov decision processes: A
  convergence time study.
\newblock \emph{Proceedings of the ACM on Measurement and Analysis of Computing
  Systems}, 2\penalty0 (1):\penalty0 1--38, 2018.

\bibitem[Zhang(2011)]{zhang2011improved}
D.~Zhang.
\newblock An improved dynamic programming decomposition approach for network
  revenue management.
\newblock \emph{Manufacturing \& Service Operations Management}, 13\penalty0
  (1):\penalty0 35--52, 2011.

\bibitem[Zhang and Adelman(2009)]{zhang2009approximate}
D.~Zhang and D.~Adelman.
\newblock An approximate dynamic programming approach to network revenue
  management with customer choice.
\newblock \emph{Transportation Science}, 43\penalty0 (3):\penalty0 381--394,
  2009.

\bibitem[Zhang et~al.(2022)Zhang, Samiedaluie, and Zhang]{zhang2022product}
R.~Zhang, S.~Samiedaluie, and D.~Zhang.
\newblock Product-based approximate linear programs for network revenue
  management.
\newblock \emph{Operations Research}, 70\penalty0 (5):\penalty0 2837--2850,
  2022.

\bibitem[Zheng and Ratliff(2020)]{zheng2020constrained}
L.~Zheng and L.~Ratliff.
\newblock Constrained upper confidence reinforcement learning.
\newblock In \emph{Learning for Dynamics and Control}, pages 620--629. PMLR,
  2020.

\end{thebibliography}

\clearpage

%
%
%

\begin{APPENDICES}
\crefalias{section}{appendix}

\section{Customer Arrival Correlation Model}\label{sec:Corremodel}

We now provide more discussions and illustrations over our approach to model the correlation of the customer arrival. We also compare with existing approaches in the literature that model correlated customer arrivals.

The transition of the state $\bs_t$ corresponds to a Markov chain. In this sense, our arrival model is analogous to the so-called \textit{Markovian-modulated} demand process, which has been extensively studied in the literature of inventory management (e.g., \citet{song1993inventory, sethi1997optimality, chen2001optimal}). The use of Markovian-modulated demand processes has been reported in \cite{simchi2004handbook} for a wider range of applications in supply chain management. There has also been a use of Markovian-modulated demand processes in the revenue management and pricing literature, see, for example, \cite{rustichini1995learning}, \cite{aviv2005partially}, and \cite{keskin2022selling}. Therefore, it is common in practice to assume that customer demand would arrive according to a Markov process, and the traditional Poisson arrival process for the NRM problem \citep{gallego1994optimal} can be viewed as a special case of the Markov arrival process.
However, the Markovian-modulated demand process studied in the previous papers all assumes that the transition between states is homogeneous over the entire horizon. In contrast, in our customer arrival model, we allow the transition probabilities to be \textit{non-homogeneous} across time, which allows us to capture more customer arrival models, especially correlated ones, that have been studied in the literature as special cases. We illustrate in the following paragraphs.

\textbf{Independent arrival model.} It is clear that if the state transition probabilities in our model are independent of the current state, i.e., $p_t(\bs, \bs')$ are common for each $\bs'\in\mathcal{S}$, then our arrival model recovers the independent customer arrival model. Moreover, the non-homogeneity of $\{p_t(\bs, \bs'), \forall \bs, \bs'\in\mathcal{S}\}$ over $t$ allows us to capture the non-stationarity of customer arrivals, as studied in \cite{ma2020approximation} and \cite{jiang2020online}. We now focus on the correlated arrival models studied in the literature.

\textbf{High-variance correlated model in \cite{bai2022fluid}.} In \cite{bai2022fluid}, the following model is adopted to capture the possible high variance in the arrival of each type of customer. A random variable $D$ is used to capture the number of customer arrivals, and a survival rate $\rho_t=P(D\geq t+1| D\geq t)$ is defined. Then, at each period $t$, conditional on $D\geq t$, one customer arrives, and customer $t$ is of type $j\in[n]$ with probability $\lambda_{j,t}$. The high-variance model can be captured by letting $\mathcal{S}=\{0, 1, \dots, n\}$, where state $j\in[m]$ denotes the customer is of type $j$, and state $0$ denotes no customer arrival. Then, the transition probabilities in our model can be defined as follows:
\[
p_t(j, j')=\rho_{t-1}\cdot\lambda_{j',t},~ p_t(j,0)=1-\rho_{t-1},\text{~and~}p_t(0,0)=1,~\forall j, j'\in[n], \forall t\in[T],
\]
to capture the high variance arrival model.

\textbf{$\INDEP$ and $\CORREL$ correlated models in \cite{aouad2022nonparametric}.} The $\INDEP$ model decides the total number of arrivals of type $j$ customers according to a distribution for each $j\in[n]$, denoted as $D_j$. The $\CORREL$ model samples the total number of arrivals $D$ from a distribution and assigns each arrival to a type $j$ with probability $p_j$, for each $j\in[n]$. Then, all customers in $\INDEP$ and $\CORREL$ arrive in a uniformly random order. We can define the state at each period to be the number of each type of customer up to that point and write out the transition probabilities accordingly. In this way, $\INDEP$ and $\CORREL$ can be expressed by our model with an exponentially large number of states.

\textbf{Correlation arrival model in \cite{truong2019prophet}.} The arrival model in \cite{truong2019prophet} is similar to ours, where they assume there is an exogenous state information $S_t$ at each period $t$ that determines the type (or type distribution) of the customer. We can think of the information $S_t$ in \cite{truong2019prophet} as the state $\bs_t$ in our model. The requirement of knowing the joint distribution of ${S_t, \forall t\in[T]}$ in \cite{truong2019prophet} is also analogous to the assumption of knowing the transition probabilities $\{p_t(\bs, \bs'), \forall \bs, \bs'\in\mathcal{S}\}$ in our model.

\section{Missing Proofs for \Cref{sec:UB}}\label{sec:pf3}

\begin{myproof}[Proof of \Cref{lem:LPupperbound}]
Plug the linear approximation \eqref{eqn:approximate} into the LP formulation of the DP policy in \eqref{lp:DP}. The objective function would become
\begin{equation}\label{eqn:042513}
\sum_{\bs\in\mathcal{S}}p_1(\bs)\cdot V_{1}(\bm{C}, \bs)=\sum_{\bs\in\mathcal{S}} p_1(\bs)\cdot \left( \theta^1(\bs)+\sum_{i\in[m]}C_i\cdot\beta^1_{i}(\bs) \right),
\end{equation}
which corresponds to the objective function \eqref{Ob:ADP} in LP \eqref{lp:ADP}. Also, constraint \eqref{const:DP1} would become
\begin{align}
\theta^t(\bs)-\sum_{\bs'\in\mathcal{S}}p_t(\bs, \bs')\cdot\theta^{t+1}(\bs')\geq& r_{j(\bs)}-\sum_{\bs'\in\mathcal{S}}p_t(\bs, \bs')\cdot\sum_{i\in[m]}a_{i,j(\bs)}\cdot \beta^{t+1}_{i}(\bs')\label{eqn:042510}\\
&+\sum_{i\in[m]}c_i\cdot \left( \sum_{\bs'\in\mathcal{S}}p_t(\bs, \bs')\cdot\beta^{t+1}_i(\bs')-\beta^t_i(\bs) \right), \forall \bc\geq\ba_{j(\bs)}, \forall t\in[T], \forall \bs\in\mathcal{S},\nonumber
\end{align}
and constraint \eqref{const:DP2} would become
\begin{equation}\label{eqn:042511}
\theta^t(\bs)-\sum_{\bs'\in\mathcal{S}}p_t(\bs, \bs')\cdot\theta^{t+1}(\bs')\geq\sum_{i\in[m]}c_i\cdot \left( \sum_{\bs'\in\mathcal{S}}p_t(\bs, \bs')\cdot\beta^{t+1}_i(\bs')-\beta^t_i(\bs) \right), \forall \bc, \forall t\in[T], \forall \bs\in\mathcal{S}.
\end{equation}
A unified way to express the constraint \eqref{eqn:042510} and \eqref{eqn:042511} would be
\begin{align}
\theta^t(\bs)-\sum_{\bs'\in\mathcal{S}}p_t(\bs, \bs')\cdot\theta^{t+1}(\bs')\geq& \left( r_{j(\bs)}-\sum_{\bs'\in\mathcal{S}}p_t(\bs, \bs')\cdot\sum_{i\in[m]}a_{i,j(\bs)}\cdot \beta^{t+1}_{i}(\bs') \right)\cdot \bI_{\{ \bc\geq\ba_{j(\bs)} \}}
 \label{eqn:042512}\\
&+\sum_{i\in[m]}c_i\cdot \left( \sum_{\bs'\in\mathcal{S}}p_t(\bs, \bs')\cdot\beta^{t+1}_i(\bs')-\beta^t_i(\bs) \right), \forall \bc, \forall t\in[T], \forall \bs\in\mathcal{S}.\nonumber
\end{align}
We now compare the constraint \eqref{eqn:042512} with the constraint \eqref{const:ADP} in LP \eqref{lp:ADP}. Clearly, for any set of variables $\{\beta^t_i(\bs), \forall t, \forall i, \forall \bs\}$, it holds that
\[\begin{aligned}
&\left[ r_{j(\bs)}-\sum_{\bs'\in\mathcal{S}}p_t(\bs, \bs')\cdot\sum_{i\in[m]}a_{i,j(\bs)}\cdot \beta^{t+1}_{i}(\bs') \right]^+ +\sum_{i\in[m]}C_i\cdot\left[ \sum_{\bs'\in\mathcal{S}}p_t(\bs, \bs')\cdot \beta^{t+1}_{i}(\bs')-\beta_{i}^t(\bs) \right]^+\\
\geq & \left( r_{j(\bs)}-\sum_{\bs'\in\mathcal{S}}p_t(\bs, \bs')\cdot\sum_{i\in[m]}a_{i,j(\bs)}\cdot \beta^{t+1}_{i}(\bs') \right)\cdot \bI_{\{ \bc\geq\ba_{j(\bs)} \}}
+\sum_{i\in[m]}c_i\cdot \left( \sum_{\bs'\in\mathcal{S}}p_t(\bs, \bs')\cdot\beta^{t+1}_i(\bs')-\beta^t_i(\bs) \right), \forall \bc\leq\bm{C}.
\end{aligned}\]
Therefore,
any set of solution $\{\theta^t(\bs), \beta^t_i(\bs), \forall t, \forall i, \forall \bs\}$ satisfying the constraint \eqref{const:ADP} would satisfy the constraint \eqref{eqn:042512}. Therefore, we know that constraint \eqref{const:ADP} in LP \eqref{lp:ADP} is stricter than the constraint \eqref{eqn:042512} above, which implies that the optimal objective value of LP \eqref{lp:ADP} is an upper bound of the optimal objective value of the LP with the objective function being \eqref{eqn:042513} and constraint being \eqref{eqn:042512}. The latter is again an upper bound of the optimal objective value of LP \eqref{lp:DP} since the linear approximation formulation \eqref{eqn:approximate} has been plugged in to restrict the range of the decision variable $V_t(\bc, \bs)$ in \eqref{lp:DP}. The above arguments show that
\[
\hat{V}^*\geq \mathbb{E}_{I\sim \mathcal{F}}[V^{\pi^*}(I)],
\]
which completes our proof.

\end{myproof}

\begin{myproof}[Proof of \Cref{prop:tightUB}]
For the high variance correlated model, an LP upper bound that has been developed in the previous literature (e.g. \cite{bai2022fluid} and \cite{aouad2022nonparametric}) can be formulated as follows.
\begin{subequations}\label{lp:UF}
\begin{align}
V^{\text{UF}}=\max&~\sum_{t\in[T]}\sum_{j\in[n]} P(\bs_t>0)\cdot r_j\cdot x_{j,t}\\  \mbox{s.t.}&~\sum_{t\in[T]}\sum_{j\in\mathcal{B}_i} x_{j,t}\leq C_i,~~\forall i\in[m] \label{const:UF1}\\
   &~x_{j,t}\leq\lambda_{j,t},~~\forall j\in[n], \forall t\in[T].\label{const:UF2}
\end{align}
\end{subequations}
where $\mathcal{B}_i$ denotes the set of customer types that require at least one unit of resource $i$ to be served. Here, the variable $x_{j,t}$ can be interpreted as the expected number of times that customer $t$ of type $j$ is served, conditional on customer $t$ arriving. Note that it has been shown in \cite{bai2022fluid} that $V^{\text{UF}}$ in \eqref{lp:UF} is asymptotically tight with respect to the optimal policy as the initial capacities $\bm{C}$ are scaled up to infinity. Then, we only need to show that the value of LP \eqref{lp:ADP} is no larger than the value of $V^{\text{UF}}$ in \eqref{lp:UF}.

We consider the dual LP of $V^{\text{UF}}$ in \eqref{lp:UF} and we use the optimal solution of the dual LP to construct a feasible solution to LP \eqref{lp:ADP} with the same objective value.
We introduce a dual variable $\mu_i$ for constraint \eqref{const:UF1}, for each $i\in[m]$. We introduce a dual variable $y_{j,t}$ for constraint \eqref{const:UF2}, for each $j\in[n]$ and each $t\in[T]$. Then, the dual LP of $V^{\text{UF}}$ in \eqref{lp:UF} can be given as follows.
\begin{subequations}\label{lp:DualUF}
\begin{align}
V^{\text{Dual}}= \min &~\sum_{t\in[T]}\sum_{j\in[n]}\lambda_{j,t}\cdot y_{j,t}+ \sum_{i\in[m]}C_i\cdot \mu_i \\
\mbox{s.t.}&~y_{j,t}\geq P(\bs_t>0)\cdot r_j-\sum_{i\in\mathcal{A}_j} \mu_i,~~\forall j\in[n], \forall t\in[T]\\
&~y_{j,t}\geq0, \mu_i\geq0, \forall i\in[m], \forall j\in[n], \forall t\in[T],
\end{align}
\end{subequations}
where $\mathcal{A}_j$ denotes the set of resources that customer of type $j\in[n]$ would require.

Under the high-variance correlated model, the state space is given as $\mathcal{S}=\{0,1,\dots, n\}$, where state $j\in[m]$ denotes the customer is of type $j$ and the state $0$ denotes there is no customer arrival. Without loss of generality, we assume there is at least one customer arrival, i.e., $P(\bs_1>0)=1$. Then, the formulation of LP \eqref{lp:ADP} can be simplified as follows.
\begin{subequations}\label{lp:ADP-HV}
\begin{align}
\min ~~&\sum_{j\in\mathcal{S}} p_1(j)\cdot \left( \theta^1(j)+\sum_{i\in[m]}C_i\cdot\beta^1_{i}(j) \right)\label{Ob:ADP-HV}\\
\mbox{s.t.}~~& \theta^t(j)-\sum_{j'\in\mathcal{S}}p_t(j, j')\cdot\theta^{t+1}(j')\geq \left[ r_{j}-\sum_{j'\in\mathcal{S}}p_t(j, j')\cdot\sum_{i\in\mathcal{A}_j} \beta^{t+1}_{i}(j') \right]^+  &\nonumber\\
&~~~~~~~~+\sum_{i\in[m]}C_i\cdot\left[ \sum_{j'\in\mathcal{S}}p_t(j, j')\cdot \beta^{t+1}_{i}(j')-\beta_{i}^t(j) \right]^+, &\forall t\in[T], \forall j\in\mathcal{S}  \label{const:ADP-HV}\\
&\theta^t(j)\geq0, \beta^t_{i}(j)\geq0, \forall i\in[m], &\forall t\in[T], \forall j\in\mathcal{S}.
\end{align}
\end{subequations}
Denote by $\{y^*_{j,t}, \mu^*_i, \forall i\in[m], \forall j\in[n], \forall t\in[T]\}$ an optimal solution to the dual LP \eqref{lp:DualUF}. We now construct a feasible solution to LP \eqref{lp:ADP-HV} with the same objective value as $V^{\text{Dual}}$. To be specific, we construct
\begin{equation}\label{eqn:043101}
\hat{\beta}^t_i(j)=\frac{\mu^*_i}{P(\bs_t>0)}, \forall i\in[m], \forall t\in[T], \forall j\in[m]\text{~~and~~}\hat{\beta}^t_i(0)=0, \forall i\in[m], \forall t\in[T],
\end{equation}
and we define $\hat{\theta}^{T+1}(j)=0$ for any $j\in\mathcal{S}$, $\hat{\theta}^t(0)=0$ for any $t\in[T]$, and we iteratively construct for $t=T, T-1, \dots, 1$ that
\begin{equation}\label{eqn:043102}
\hat{\theta}^t(j)-\sum_{j'\in\mathcal{S}}p_t(j, j')\cdot\hat{\theta}^{t+1}(j')= \frac{y^*_{j,t}}{P(\bs_t>0)},~~\forall j\in[n], \forall t\in[T].
\end{equation}

We first show that the solution constructed in \eqref{eqn:043101} and \eqref{eqn:043102} is feasible to LP \eqref{lp:ADP-HV}. It is clear to see that
\[
\sum_{j'\in\mathcal{S}}p_t(j, j')\cdot \hat{\beta}^{t+1}_{i}(j')-\hat{\beta}_{i}^t(j)=0,~~\forall j\in[n], \forall t\in[T]
\]
and
\[
r_{j}-\sum_{j'\in\mathcal{S}}p_t(j, j')\cdot\sum_{i\in\mathcal{A}_j} \hat{\beta}^{t+1}_{i}(j')=r_j-\frac{1}{P(\bs_t>0)}\cdot\sum_{i\in\mathcal{A}_j} \mu^*_i.
\]
Therefore, from the non-negativity of $y^*_{j,t}$, we know that for each $j\in[n]$,
\[\begin{aligned}
\hat{\theta}^t(j)-\sum_{j'\in\mathcal{S}}p_t(j, j')\cdot\hat{\theta}^{t+1}(j')=&\frac{y^*_{j,t}}{P(\bs_t>0)}\geq\left[r_j-\frac{1}{P(\bs_t>0)}\cdot\sum_{i\in\mathcal{A}_j}\mu^*_i\right]^+\\
=&\left[ r_{j}-\sum_{j'\in\mathcal{S}}p_t(j, j')\cdot\sum_{i\in\mathcal{A}_j} \hat{\beta}^{t+1}_{i}(j') \right]^+\\
&+\sum_{i\in[m]}C_i\cdot\left[ \sum_{j'\in\mathcal{S}}p_t(j, j')\cdot \hat{\beta}^{t+1}_{i}(j')-\beta_{i}^t(j) \right]^+
\end{aligned}\]
which justifies that constraint \eqref{const:ADP-HV} is satisfied. Therefore, we know that the solution constructed in \eqref{eqn:043101} and \eqref{eqn:043102} is feasible to LP \eqref{lp:ADP-HV}.

We now consider the objective value. From the construction \eqref{eqn:043102}, we have
\[
\sum_{j\in\mathcal{S}}p_1(j)\cdot \hat{\theta}^1(j)=\sum_{t\in[T]}\sum_{j\in[n]} P(\bs_t=j)\cdot \frac{y^*_{j,t}}{P(\bs_t>0)}
\]
We further note that $P(\bs_t=j)=P(\bs_t>0)\cdot \lambda_{j,t}$ for each $j\in[n]$ and each $t\in[T]$. We have
\begin{equation}\label{eqn:043103}
\sum_{j\in\mathcal{S}}p_1(j)\cdot \hat{\theta}^1(j)=\sum_{t\in[T]}\sum_{j\in[n]}\lambda_{j,t}\cdot y^*_{j,t}.
\end{equation}
On the other hand, it is clear to see that
\begin{equation}\label{eqn:043104}
\sum_{j\in\mathcal{S}}p_1(j)\cdot\sum_{i\in[m]}C_i\cdot\hat{\beta}^1_i(j)=\sum_{i\in[m]}C_i\cdot\mu^*_i.
\end{equation}
Combining \eqref{eqn:043103} and \eqref{eqn:043104}, we know that
\[
\sum_{j\in\mathcal{S}} p_1(j)\cdot \left( \hat{\theta}^1(j)+\sum_{i\in[m]}C_i\cdot\hat{\beta}^1_{i}(j) \right)=\sum_{t\in[T]}\sum_{j\in[n]}\lambda_{j,t}\cdot y^*_{j,t}+\sum_{i\in[m]}C_i\cdot\mu^*_i=V^{\text{Dual}}
\]
which completes our proof.

\end{myproof}

\section{Missing Proofs for \Cref{sec:Analysis}}\label{sec:pf5}

\begin{myproof}[Proof of \Cref{lem:lowerbound}]
We decompose the total reward collected by our \Cref{alg:BMP}, denoted by $\pi$, into two parts based on the bid price \eqref{eqn:042901}, and we have
\begin{equation}\label{eqn:042906}
\begin{aligned}
\sum_{t\in[T]}\bI_{\{x^{\pi}_t(\bs_t)=1\}}\cdot r_{j(\bs_t)}=&\underbrace{\sum_{t\in[T]} \bI_{\{x^{\pi}_t(\bs_t)=1\}}\cdot \left[ r_{j(\bs_t)}-\sum_{\bs'\in\mathcal{S}}p_t(\bs_t, \bs')\cdot\sum_{i\in\mathcal{A}_{j(\bs_t)}}\frac{1}{C_i}\cdot\sum_{j'\in\mathcal{B}_i}\nu^{t+1}_{j'}(\bs') \right]}_{\text{I}}\\
&+\underbrace{\sum_{t\in[T]} \bI_{\{x^{\pi}_t(\bs_t)=1\}}\cdot \left( \sum_{\bs'\in\mathcal{S}}p_t(\bs_t, \bs')\cdot\sum_{i\in\mathcal{A}_{j(\bs_t)}}\frac{1}{C_i}\cdot\sum_{j'\in\mathcal{B}_i}\nu^{t+1}_{j'}(\bs') \right)}_{\text{II}}.
\end{aligned}
\end{equation}
We now analyze term I and term II separately.

\noindent \textbf{Bound term I.} From the decision rule \eqref{eqn:042904}, we have
\begin{align}
\text{term I}&=\sum_{t\in[T]} \bI_{\{x^{\pi}_t(\bs_t)=1\}}\cdot \left[ r_{j(\bs_t)}-\sum_{\bs'\in\mathcal{S}}p_t(\bs_t, \bs')\cdot\sum_{i\in\mathcal{A}_{j(\bs_t)}}\frac{1}{C_i}\cdot\sum_{j'\in\mathcal{B}_i}\nu^{t+1}_{j'}(\bs') \right]\nonumber\\
&=\sum_{t\in[T]}\bI_{\{\bc_t\geq \ba_{j(\bs_t)}\}}\cdot \left[ r_{j(\bs_t)}-\sum_{\bs'\in\mathcal{S}}p_t(\bs_t, \bs')\cdot\sum_{i\in\mathcal{A}_{j(\bs_t)}}\frac{1}{C_i}\cdot\sum_{j'\in\mathcal{B}_i}\nu^{t+1}_{j'}(\bs') \right]^+.\label{eqn:042907}
\end{align}

\noindent \textbf{Bound term II.} From the bid price computing rule \eqref{eqn:042202}, for any $t\in[T]$, $j\in[n]$, $\bs\in\mathcal{S}$, we know that
\[
\nu^t_j(\bs)=\sum_{t'=t}^{T}\sum_{\bs'\in\mathcal{S}} P( \bs_{t'}=\bs'|\bs_t=\bs)\cdot\bI_{\{j=j(\bs')\}}\cdot\left[r_j-\sum_{\bs''\in\mathcal{S}}p_{t'}(\bs', \bs'')\cdot\sum_{i\in\mathcal{A}_j}\frac{1}{C_i}\cdot\sum_{j'\in\mathcal{B}_i}\nu^{t'+1}_{j'}(\bs'')\right]^+,
\]
where we use $P(\bs_{t'}=\bs'|\bs_t=\bs)$ to denote the probability that conditional on $\bs_t=\bs$, $\bs_{t'}$ is realized as $\bs'$. For each period $t\in[T]$, we also denote by $\bm{b}_t=\bm{C}-\bc_t$ the units of resources consumed at the beginning of period $t$. Then, by switching the sums, we have
\[\begin{aligned}
&\mathbb{E}[\text{term II}]\\
=&\mathbb{E}\left[ \sum_{t\in[T]} \bI_{\{x^{\pi}_t(\bs_t)=1\}}\cdot \left( \sum_{\bs'\in\mathcal{S}}p_t(\bs_t, \bs')\cdot\sum_{i\in\mathcal{A}_{j(\bs_t)}}\frac{1}{C_i}\cdot\sum_{j'\in\mathcal{B}_i}\nu^{t+1}_{j'}(\bs') \right) \right]\\
=&\mathbb{E}\left[ \sum_{t\in[T]}\left(\left[ r_{j(\bs_t)}-\sum_{\bs'\in\mathcal{S}}p_t(\bs_t, \bs')\cdot\sum_{i\in\mathcal{A}_{j(\bs_t)}}\frac{1}{C_i}\cdot\sum_{j'\in\mathcal{B}_i}\nu^{t+1}_{j'}(\bs') \right]^+\cdot \sum_{i\in\mathcal{A}_{j(\bs_t)}}\frac{1}{C_i}\cdot\sum_{t'=1}^{t-1}\bI_{\{x^{\pi}_{t'}(\bs_{t'})=1\text{~and~}i\in\mathcal{A}_{j(\bs_{t'})}\}} \right)\right]\\
=& \sum_{t\in[T]}\sum_{\bs\in\mathcal{S}}P(\bs_t=\bs)\cdot\left(\left[ r_{j(\bs)}-\sum_{\bs'\in\mathcal{S}}p_t(\bs, \bs')\cdot\sum_{i\in\mathcal{A}_{j(\bs_t)}}\frac{1}{C_i}\cdot\sum_{j'\in\mathcal{B}_i}\nu^{t+1}_{j'}(\bs') \right]^+\cdot \sum_{i\in\mathcal{A}_{j(\bs)}}\frac{\mathbb{E}[b_{i,t}|\bs_t=\bs]}{C_i}\right).
\end{aligned}\]
We apply Markov's inequality to show that for each $i\in[m]$, it holds
\[
\frac{\mathbb{E}[b_{i,t}|\bs_t=\bs]}{C_i}\geq P(b_{i,t}\geq C_i|\bs_t=\bs)=P(c_{i,t}\leq 0|\bs_t=\bs).
\]
Then, by the union bound, we have
\[
\sum_{i\in\mathcal{A}_{j(\bs)}}\frac{\mathbb{E}[b_{i,t}|\bs_t=\bs]}{C_i}\geq \sum_{i\in\mathcal{A}_{j(\bs)}}P(c_{i,t}\leq 0|\bs_t=\bs) \geq P\left(\exists i\in\mathcal{A}_{j(\bs)}: c_{i,t}< a_{i,j(\bs)}|\bs_t=\bs\right)=1-P(\bc_t\geq \ba_{j(\bs)}|\bs_t=\bs).
\]
Therefore, we know that
\begin{equation}\label{eqn:043001}
\begin{aligned}
&\mathbb{E}[\text{term II}]\\
=&\sum_{t\in[T]}\sum_{\bs\in\mathcal{S}}P(\bs_t=\bs)\cdot\left(\left[ r_{j(\bs)}-\sum_{\bs'\in\mathcal{S}}p_t(\bs, \bs')\cdot\sum_{i\in\mathcal{A}_{j(\bs_t)}}\frac{1}{C_i}\cdot\sum_{j'\in\mathcal{B}_i}\nu^{t+1}_{j'}(\bs') \right]^+\cdot \sum_{i\in\mathcal{A}_{j(\bs)}}\frac{\mathbb{E}[b_{i,t}|\bs_t=\bs]}{C_i}\right)\\
\geq& \sum_{t\in[T]}\sum_{\bs\in\mathcal{S}}P(\bs_t=\bs)\cdot\left(\left[ r_{j(\bs)}-\sum_{\bs'\in\mathcal{S}}p_t(\bs, \bs')\cdot\sum_{i\in\mathcal{A}_{j(\bs_t)}}\frac{1}{C_i}\cdot\sum_{j'\in\mathcal{B}_i}\nu^{t+1}_{j'}(\bs') \right]^+\cdot \left( 1-P(\bc_t\geq \ba_{j(\bs)}|\bs_t=\bs) \right)   \right).
\end{aligned}
\end{equation}
On the other hand, from \eqref{eqn:042907}, we know
\begin{equation}\label{eqn:043002}
    \mathbb{E}[\text{term I}]=\sum_{t\in[T]}\sum_{\bs\in\mathcal{S}}P(\bs_t=\bs)\cdot\left(\left[ r_{j(\bs)}-\sum_{\bs'\in\mathcal{S}}p_t(\bs, \bs')\cdot\sum_{i\in\mathcal{A}_{j(\bs_t)}}\frac{1}{C_i}\cdot\sum_{j'\in\mathcal{B}_i}\nu^{t+1}_{j'}(\bs') \right]^+\cdot P(\bc_t\geq \ba_{j(\bs)}|\bs_t=\bs)   \right).
\end{equation}
Combining \eqref{eqn:043001}, \eqref{eqn:043002}, and \eqref{eqn:042906}, we have
\begin{equation}\label{eqn:043003}
\mathbb{E}\left[\sum_{t\in[T]}\bI_{\{x^{\pi}_t(\bs_t)=1\}}\cdot r_{j(\bs_t)}\right]\geq\sum_{t\in[T]}\sum_{\bs\in\mathcal{S}}P(\bs_t=\bs)\cdot\left[ r_{j(\bs)}-\sum_{\bs'\in\mathcal{S}}p_t(\bs, \bs')\cdot\sum_{i\in\mathcal{A}_{j(\bs_t)}}\frac{1}{C_i}\cdot\sum_{j'\in\mathcal{B}_i}\nu^{t+1}_{j'}(\bs') \right]^+.
\end{equation}
Finally, from the bid price computing rule \eqref{eqn:042202} and by noting $\nu^{T+1}_j(\bs)=0$ for any $j\in[n]$ and $\bs\in\mathcal{S}$, we also have
\begin{equation}\label{eqn:043004}
 \sum_{\bs\in\mathcal{S}}p_1(\bs)\cdot \sum_{j\in[n]}\nu^1_j(\bs)=  \sum_{t\in[T]}\sum_{\bs\in\mathcal{S}}P(\bs_t=\bs)\cdot\left[ r_{j(\bs)}-\sum_{\bs'\in\mathcal{S}}p_t(\bs, \bs')\cdot\sum_{i\in\mathcal{A}_{j(\bs_t)}}\frac{1}{C_i}\cdot\sum_{j'\in\mathcal{B}_i}\nu^{t+1}_{j'}(\bs') \right]^+.
\end{equation}
Comparing \eqref{eqn:043003} and \eqref{eqn:043004}, we have
\[
\mathbb{E}\left[\sum_{t\in[T]}\bI_{\{x^{\pi}_t(\bs_t)=1\}}\cdot r_{j(\bs_t)}\right]\geq \sum_{\bs\in\mathcal{S}}p_1(\bs)\cdot \sum_{j\in[n]}\nu^1_j(\bs)
\]
which completes our proof.

\end{myproof}

\begin{myproof}[Alternative proof of \Cref{lem:lowerbound}]
Denote by $\pi$ our \Cref{alg:BMP} and denote by $H^{\pi}_{t}(\bc, \bs)$ the total expected reward collected by the policy $\pi$ from period $t$ to period $T$, given the remaining capacity at period $t$ is $\bc$ and the state of period $t$ is $\bs$. It is clear to see that $H^{\pi}_{t}(\bc, \bs)$ admits the following backward induction, which is similar to the backward induction of the DP in \eqref{eqn:backward},
\begin{equation}\label{eqn:Policybackward}
H^{\pi}_{t}(\bc, \bs)= r_{j(\bs)}\cdot x^{\pi}_t(\bs)+\sum_{\bs'\in\mathcal{S}}p_t(\bs, \bs')\cdot H^{\pi}_{t+1}(\bc-\ba_{j(\bs)}\cdot x^{\pi}_t(\bs), \bs')
\end{equation}
where $x^{\pi}_t(\bs)\in\{0,1\}$ indicates whether customer $t$ has been served or not by the policy $\pi$, given the current state is realized as $\bs$. The key step of our proof is to utilize a basis function $\psi_j(\bc)$ for each $j\in[n]$ such that
\begin{equation}\label{eqn:042501}
H^{\pi}_{t}(\bc, \bs) \geq \sum_{j\in[n]}\psi_{j}(\bc)\cdot \nu_{j}^t(\bs)
\end{equation}
is satisfied for any possible $\bc$, any $t\in[T]$, and any $\bs\in\mathcal{S}$. To be more concrete, similar to \cite{ma2020approximation}, we utilize the basis functions constructed as follows,
\begin{equation}\label{eqn:042502}
\psi_j(\bc)=\prod_{i\in\mathcal{A}_j}\frac{c_i}{C_i}.
\end{equation}
The basis function defined in \eqref{eqn:042502} preserves many nice properties, which will be helpful for us to prove \eqref{eqn:042501}. Finally, note that $\psi_j(\bm{C})=1$ for each $j\in[n]$. We prove that $H^{\pi}_1(\bm{C}, \bs)$ is lower bounded by $\sum_{j\in[n]}\nu^1_j(\bs)$, for any $\bs\in\mathcal{S}$.

Let the basis function $\psi_j(\bc)$ be defined in \eqref{eqn:042502}. It is sufficient to prove \eqref{eqn:042501}.
It is clear to see that
\[
H^{\pi}_{T+1}(\bc, \bs)=0\geq 0=\sum_{j\in[n]}\psi_{j}(\bc)\cdot \nu_{j}^{T+1}(\bs).
\]
We prove \eqref{eqn:042501} inductively for $t=T, T-1, \dots, 1$. Suppose that \eqref{eqn:042501} is satisfied for $t+1$ and any $\bc$, any $\bs\in\mathcal{S}$. We now prove \eqref{eqn:042501} would also be satisfied for $t$. We note that
\[\begin{aligned}
H^{\pi}_{t}(\bc, \bs)&= r_{j(\bs)}\cdot y^{\pi}_t(\bs)+\sum_{\bs'\in\mathcal{S}}p_t(\bs, \bs')\cdot H^{\pi}_{t+1}(\bc-\ba_{j(\bs)}\cdot y^{\pi}_t(\bs), \bs')\\
&\geq r_{j(\bs)}\cdot y^{\pi}_t(\bs)+\sum_{\bs'\in\mathcal{S}}p_t(\bs, \bs')\cdot \left(\sum_{j\in[n]}\nu^{t+1}_{j}(\bs')\cdot\psi_j\left(\bc-\ba_{j(\bs)}\cdot y^{\pi}_t(\bs)\right) \right)
\end{aligned}\]
holds for any $\bc$ and any $\bs\in\mathcal{S}$. Therefore, in order to prove $H^{\pi}_{t}(\bc, \bs)\geq\sum_{j\in[n]}\nu^t_j(\bs)\cdot\psi_j(\bc)$, it is sufficient to prove that the inequality
\[
r_{j(\bs)}\cdot y^{\pi}_t(\bs)+\sum_{\bs'\in\mathcal{S}}p_t(\bs, \bs')\cdot \left(\sum_{j\in[n]}\nu^{t+1}_{j}(\bs')\cdot\psi_j\left(\bc-\ba_{j(\bs)}\cdot y^{\pi}_t(\bs)\right) \right)\geq \sum_{j\in[n]}\nu^t_j(\bs)\cdot\psi_j(\bc)
\]
holds for any $\bc$ and any $\bs\in\mathcal{S}$. The above inequality is equivalent to
\begin{equation}\label{eqn:042503}
\begin{aligned}
&\sum_{j\in[n]}\left(\psi_{j}(\bc)\cdot\nu^t_{j}(\bs)-\sum_{\bs'\in\mathcal{S}}p_t(\bs, \bs')\cdot\psi_{j}(\bc)\cdot\nu^{t+1}_{j}(\bs')  \right)=\psi_{j(\bs)}(\bc)\cdot \left[r_{j(\bs)}-\sum_{\bs'\in\mathcal{S}}p_t(\bs, \bs')\cdot\sum_{i\in\mathcal{A}_{j(\bs)}}\frac{1}{C_i}\cdot\sum_{j'\in\mathcal{B}_i}\nu^{t+1}_{j'}(\bs') \right]^+\\
&\leq r_{j(\bs)}\cdot y^{\pi}_t(\bs)-\sum_{\bs'\in\mathcal{S}}p_t(\bs, \bs')\cdot \left(\sum_{j\in[n]}\nu^{t+1}_{j}(\bs')\cdot\left(\psi_{j}(\bc)-\psi_{j}(\bc-\ba_{j(\bs)}\cdot y^{\pi}_t(\bs)) \right)\right).
\end{aligned}
\end{equation}
where the equality follows from the definition of $\nu^t_j(\bs)$ in \eqref{eqn:042202}.
We consider two scenarios separately based on the value of $\bc$.

\noindent Scenario i: there exists an $i\in[m]$ such that $c_i<a_{i, j(\bs)}$. Then, following \Cref{alg:BMP}, we have $y^{\pi}_t(\bs)=0$, which implies that the right hand side of inequality \eqref{eqn:042503} equals $0$. Also, note that $\psi_{j(\bs)}(\bc)=0$ and thus the left hand side of inequality \eqref{eqn:042503} also equals $0$. We prove inequality \eqref{eqn:042503} indeed holds for this scenario.

\noindent Scenario ii: for all $i\in[n]$ we have $c_i\geq a_{i,j(\bs)}$.
On one hand, by noting that $\psi_j(\bc)\leq 1$ for any $\bc\leq\bm{C}$ and any $j\in[n]$, we have
\begin{equation}\label{eqn:042504}
\begin{aligned}
\psi_{j(\bs)}(\bc)\cdot \left[r_{j(\bs)}-\sum_{\bs'\in\mathcal{S}}p_t(\bs, \bs')\cdot\sum_{i\in\mathcal{A}_{j(\bs)}}\frac{1}{C_i}\cdot\sum_{j'\in\mathcal{B}_i}\nu^{t+1}_{j'}(\bs') \right]^+
\leq\left[r_{j(\bs)}-\sum_{\bs'\in\mathcal{S}}p_t(\bs, \bs')\cdot\sum_{i\in\mathcal{A}_{j(\bs)}}\frac{1}{C_i}\cdot\sum_{j'\in\mathcal{B}_i}\nu^{t+1}_{j'}(\bs') \right]^+.
\end{aligned}
\end{equation}
Following the definition of $y^{\pi}_t(\bs)$ in \Cref{alg:BMP}, we know that
\[
\left[r_{j(\bs)}-\sum_{\bs'\in\mathcal{S}}p_t(\bs, \bs')\cdot\sum_{i\in\mathcal{A}_{j(\bs)}}\frac{1}{C_i}\cdot\sum_{j'\in\mathcal{B}_i}\nu^{t+1}_{j'}(\bs') \right]^+=y^{\pi}_t(\bs)\cdot \left(r_{j(\bs)}-\sum_{\bs'\in\mathcal{S}}p_t(\bs, \bs')\cdot\sum_{i\in\mathcal{A}_{j(\bs)}}\frac{1}{C_i}\cdot\sum_{j'\in\mathcal{B}_i}\nu^{t+1}_{j'}(\bs') \right).
\]
Therefore, in order to prove \eqref{eqn:042503}, it is sufficient to show that
\begin{equation}\label{eqn:042505}
\sum_{\bs'\in\mathcal{S}}p_t(\bs, \bs')\cdot \left(\sum_{j\in[n]}\nu^{t+1}_{j}(\bs')\cdot\left(\psi_{j}(\bc)-\psi_{j}(\bc-\ba_{j(\bs)}\cdot y^{\pi}_t(\bs)) \right)\right)\leq y^{\pi}_t(\bs)\cdot \left(\sum_{\bs'\in\mathcal{S}}p_t(\bs, \bs')\cdot\sum_{i\in\mathcal{A}_{j(\bs)}}\frac{1}{C_i}\cdot\sum_{j'\in\mathcal{B}_i}\nu^{t+1}_{j'}(\bs') \right).
\end{equation}
If $y^{\pi}_t(\bs)=0$, it is clear to see that both sides of inequality \eqref{eqn:042505} equal $0$ and inequality \eqref{eqn:042505} thus hold. We now suppose $y^{\pi}_t(\bs)=1$. We have the following claim, which corresponds to the \textit{maximum scaled incremental contribution} requirement in \cite{ma2020approximation} and can be directly verified following the definition of $\psi_j(\cdot)$ in \eqref{eqn:042502}.
\begin{claim}\label{claim:basis}
Let the basis function $\psi_j(\cdot)$ be defined in \eqref{eqn:042502} for each $j\in[n]$. Then, for any $j\in[n]$ and any $i\in[m]$ such that $a_{i,j}=0$, we have
\[
\psi_j(\bc+\bm{e}_i)-\psi_j(\bc)=0.
\]
where $\bm{e}_i\in\mathbb{R}^m$ denotes a vector with the $i$-th element being $1$ and all other elements being $0$. For any $j\in[n]$ and any $i\in[m]$ such that $a_{i,j}=1$, we have
\[
\psi_j(\bc+\bm{e}_i)-\psi_j(\bc) \leq \frac{1}{C_i}.
\]
\end{claim}
Following \Cref{claim:basis}, for any $j\in[n]$, we have
\[
\psi_{j}(\bc)-\psi_{j}(\bc-\ba_{j(\bs)})\leq \sum_{i: i\in\mathcal{A}_j\cap\mathcal{A}_{j(\bs)} } \frac{1}{C_i}.
\]
Therefore, for any $\bs'\in\mathcal{S}$, we have
\[
\sum_{j\in[n]}\nu^{t+1}_{j}(\bs')\cdot\left(\psi_{j}(\bc)-\psi_{j}(\bc-\ba_{j(\bs)}) \right)\leq \sum_{j\in[n]}\nu^{t+1}_{j}(\bs')\cdot \sum_{i: i\in\mathcal{A}_j\cap\mathcal{A}_{j(\bs)} } \frac{1}{C_i}\leq \sum_{i\in\mathcal{A}_{j(\bs)}}\frac{1}{C_i}\cdot\sum_{j'\in\mathcal{B}_i}\nu^{t+1}_{j'}(\bs'),
\]
which completes our proof of \eqref{eqn:042505}.

From the steps above, we show that \eqref{eqn:042501} being satisfied for $t+1$ and any $\bc$, any $\bs\in\mathcal{S}$, would imply \eqref{eqn:042501} being satisfied for $t$ and any $\bc$, any $\bs\in\mathcal{S}$. Follwoing the induction argument, we prove \eqref{eqn:042501} is satisfied for any $t\in[T]$, any $\bc$, and any $\bs\in\mathcal{S}$, which completes our proof.

\end{myproof}

\begin{myproof}[Proof of \Cref{lem:upperbound}]
We first show that $\{\hat{\beta}^t_i(\bs), \hat{\theta}^t(\bs), \forall i\in[m], \forall t\in[T], \forall \bs\in\mathcal{S}\}$ defined in \eqref{eqn:042301} and \eqref{eqn:042302} is feasible to LP \eqref{lp:ADP}. Note that from the definition in \eqref{eqn:042202}, we must have $\nu^t_j(\bs)\geq\sum_{\bs'\in\mathcal{S}}p_t(\bs, \bs')\cdot\nu^{t+1}_j(\bs')$ for any $j\in[n]$, $t\in[T]$ and $\bs\in\mathcal{S}$, which implies that
\[
\hat{\beta}^t_i(\bs)\geq\sum_{\bs'\in\mathcal{S}}p_t(\bs, \bs')\cdot\hat{\beta}^{t+1}_i(\bs').
\]
Therefore, in order for the constraint \eqref{const:ADP} to be satisfied, it is sufficient to show that
\[
 \hat{\theta}^t(\bs)-\sum_{\bs'\in\mathcal{S}}p_t(\bs, \bs')\cdot\hat{\theta}^{t+1}(\bs')\geq \left[ r_{j(\bs)}-\sum_{\bs'\in\mathcal{S}}p_t(\bs, \bs')\cdot\sum_{i\in[m]}a_{i,j(\bs)}\cdot \hat{\beta}^{t+1}_{i}(\bs') \right]^+.
\]
Note that
\[\begin{aligned}
\hat{\theta}^t(\bs)-\sum_{\bs'\in\mathcal{S}}p_t(\bs, \bs')\cdot\hat{\theta}^{t+1}(\bs')&=\sum_{j\in[n]}\nu^t_j(\bs)-\sum_{\bs'\in\mathcal{S}}p_t(\bs, \bs')\cdot\sum_{j\in[n]}\nu^t_j(\bs')\\
&=\left[r_{j(\bs)}-\sum_{\bs'\in\mathcal{S}}p_t(\bs, \bs')\cdot\sum_{i\in\mathcal{A}_{j(\bs)}}\frac{1}{C_i}\cdot\sum_{j'\in\mathcal{B}_i}\nu^{t+1}_{j'}(\bs') \right]^+\\
&=\left[r_{j(\bs)}-\sum_{\bs'\in\mathcal{S}}p_t(\bs, \bs')\cdot\sum_{i\in\mathcal{A}_{j(\bs)}}\beta^{t+1}_{i}(\bs') \right]^+\\
&=\left[ r_{j(\bs)}-\sum_{\bs'\in\mathcal{S}}p_t(\bs, \bs')\cdot\sum_{i\in[m]}a_{i,j(\bs)}\cdot \hat{\beta}^{t+1}_{i}(\bs') \right]^+.
\end{aligned}\]
Therefore, we show that $\{\hat{\beta}^t_i(\bs), \hat{\theta}^t(\bs), \forall i\in[m], \forall t\in[T], \forall \bs\in\mathcal{S}\}$ defined in \eqref{eqn:042301} and \eqref{eqn:042302} is feasible to LP \eqref{lp:ADP}.

The objective value of LP \eqref{lp:ADP} under the feasible solution $\{\hat{\beta}^t_i(\bs), \hat{\theta}^t(\bs), \forall i\in[m], \forall t\in[T], \forall \bs\in\mathcal{S}\}$ can be bounded as follows.
\[\begin{aligned}
\sum_{\bs\in\mathcal{S}} p_1(\bs)\cdot \left( \hat{\theta}^1(\bs)+\sum_{i\in[m]}C_i\cdot\hat{\beta}^1_{i}(\bs) \right)&=\sum_{\bs\in\mathcal{S}} p_1(\bs)\cdot \sum_{j\in[n]}\nu^1_j(\bs)+\sum_{\bs\in\mathcal{S}}p_1(\bs)\cdot \sum_{i\in[m]}\sum_{j\in\mathcal{B}_i} \nu^t_j(\bs) \\
&=\sum_{\bs\in\mathcal{S}} p_1(\bs)\cdot \sum_{j\in[n]}\nu^1_j(\bs)+\sum_{\bs\in\mathcal{S}}p_1(\bs)\cdot \sum_{j\in[n]}\sum_{i\in\mathcal{A}_j} \nu^t_j(\bs)\\
&\leq (1+L)\cdot \sum_{\bs\in\mathcal{S}} p_1(\bs)\cdot \sum_{j\in[n]}\nu^1_j(\bs)
\end{aligned}\]
where it follows from definition \eqref{eqn:042203} that the number of elements in the set $\mathcal{A}_j$ is upper bounded by $L$ for any $j\in[n]$. Our proof is thus completed.

\end{myproof}

\section{Missing Proofs for \Cref{sec:Choicemodel}}\label{sec:pf6}

\begin{myproof}[Proof of \Cref{lem:AssortUB}]
Similar to the LP formulation \eqref{lp:DP}, we obtain the LP formulation of the DP \eqref{eqn:Assortbackward} for the assortment setting as follows.
\begin{subequations}\label{lp:AssortDP}
\begin{align}
\min ~~& \sum_{\bs\in\mathcal{S}}p_1(\bs)\cdot V_{1}(\bm{C}, \bs) \\
\mbox{s.t.} ~~& V_{t}(\bc,\bs)\geq \sum_{j\in A}\phi_j(A, \bs)\cdot\left(r_{j}+\sum_{\bs'\in\mathcal{S}}p_t(\bs, \bs')\cdot V^*_{t+1}(\bc-\ba_{j}, \bs') \right), \forall A\in\mathcal{F}(\bc), \forall \bc, \forall t\in[T], \forall \bs\in\mathcal{S} \label{const:AssortDP1}\\
& V_{t}(\bc,\bs)\geq 0, \forall \bc, \forall t\in[T], \forall \bs\in\mathcal{S}.
\end{align}
\end{subequations}
Plug the linear approximation \eqref{eqn:approximate} into the LP formulation of the DP policy in \eqref{lp:AssortDP}. The objective function would become
\begin{equation}\label{eqn:052513}
\sum_{\bs\in\mathcal{S}}p_1(\bs)\cdot V_{1}(\bm{C}, \bs)=\sum_{\bs\in\mathcal{S}} p_1(\bs)\cdot \left( \theta^1(\bs)+\sum_{i\in[m]}C_i\cdot\beta^1_{i}(\bs) \right),
\end{equation}
which corresponds to the objective function \eqref{Ob:AssortADP} in LP \eqref{lp:AssortADP}. Also, constraint \eqref{const:AssortDP1} would become
\begin{align}
\theta^t(\bs)-\sum_{\bs'\in\mathcal{S}}p_t(\bs, \bs')\cdot\theta^{t+1}(\bs')\geq& \sum_{j\in A}\phi_j(A, \bs)\cdot\left( r_{j}-\sum_{\bs'\in\mathcal{S}}p_t(\bs, \bs')\cdot\sum_{i\in[m]}a_{i,j}\cdot \beta^{t+1}_{i}(\bs')\right)\label{eqn:052510}\\
&+\sum_{i\in[m]}c_i\cdot \left( \sum_{\bs'\in\mathcal{S}}p_t(\bs, \bs')\cdot\beta^{t+1}_i(\bs')-\beta^t_i(\bs) \right), \forall A\in\mathcal{F}(\bc), \forall \bc, \forall t\in[T], \forall \bs\in\mathcal{S}.\nonumber
\end{align}
We now compare the constraint \eqref{eqn:052510} with the constraint \eqref{const:AssortADP} in LP \eqref{lp:AssortADP}. Clearly, for any set of variables $\{\beta^t_i(\bs), \forall t, \forall i, \forall \bs\}$, it holds that
\[\begin{aligned}
&\sum_{j\in A}\phi_j(A, \bs)\cdot\left[ r_{j}-\sum_{\bs'\in\mathcal{S}}p_t(\bs, \bs')\cdot\sum_{i\in[m]}a_{i,j}\cdot \beta^{t+1}_{i}(\bs') \right]^+ +\sum_{i\in[m]}C_i\cdot\left[ \sum_{\bs'\in\mathcal{S}}p_t(\bs, \bs')\cdot \beta^{t+1}_{i}(\bs')-\beta_{i}^t(\bs) \right]^+\\
\geq & \sum_{j\in A}\phi_j(A, \bs)\cdot \left( r_{j}-\sum_{\bs'\in\mathcal{S}}p_t(\bs, \bs')\cdot\sum_{i\in[m]}a_{i,j}\cdot \beta^{t+1}_{i}(\bs') \right)
+\sum_{i\in[m]}c_i\cdot \left( \sum_{\bs'\in\mathcal{S}}p_t(\bs, \bs')\cdot\beta^{t+1}_i(\bs')-\beta^t_i(\bs) \right),
\end{aligned}\]
for all $A\in\mathcal{F}(\bc)$, all $\bc\leq\bm{C}$, all state $\bs\in\mathcal{S}$ and all period $t\in[T]$.
Therefore,
any set of solution $\{\theta^t(\bs), \beta^t_i(\bs), \forall t, \forall i, \forall \bs\}$ satisfying the constraint \eqref{const:AssortADP} would satisfy the constraint \eqref{eqn:052510}. Therefore, we know that constraint \eqref{const:AssortADP} in LP \eqref{lp:AssortADP} is stricter than the constraint \eqref{eqn:052510} above, which implies that the optimal objective value of LP \eqref{lp:AssortADP} is an upper bound of the optimal objective value of the LP with the objective function being \eqref{eqn:052513} and constraint being \eqref{eqn:052510}. The latter is again an upper bound of the optimal objective value of LP \eqref{lp:AssortDP} since the linear approximation formulation \eqref{eqn:approximate} has been plugged in to restrict the range of the decision variable $V_t(\bc, \bs)$ in \eqref{lp:AssortDP}. The above arguments show that
\[
\text{LP~}\eqref{lp:AssortADP}\geq\sum_{\bs\in\mathcal{S}}p_1(\bs)\cdot V^*_1(\bm{C}, \bs),
\]
which completes our proof.

\end{myproof}

\begin{myproof}[Proof of \Cref{thm:assortCR}]
In order to prove \Cref{thm:assortCR}, we first prove the following two lemmas. The first lemma is stated below, which is analogous to \Cref{lem:lowerbound}.
\begin{lemma}\label{lem:assortlowerbound}
The expected total reward collected by our \Cref{alg:assortBMP} is lower bounded by
\[
\sum_{\bs\in\mathcal{S}}p_1(\bs)\cdot \sum_{j\in[n]}\nu^1_j(\bs),
\]
with $\nu^1_j(\bs)$ defined in \eqref{eqn:052202}.
\end{lemma}
We also prove the following lemma, which is analogous to \Cref{lem:upperbound}.
\begin{lemma}\label{lem:assortupperbound}
It holds that
\[
\text{LP~}\eqref{lp:AssortADP}\leq(1+L)\cdot \sum_{\bs\in\mathcal{S}} p_1(\bs)\cdot\sum_{j\in[n]}\nu^1_j(\bs)
\]
with $L$ defined in \eqref{eqn:042203} and $\nu^1_j(\bs)$ defined in \eqref{eqn:052202}.
\end{lemma}
The formal proofs of \Cref{lem:assortlowerbound} and \Cref{lem:assortupperbound} are included at the end of this proof. We are now ready to prove \Cref{thm:assortCR}.

From \Cref{lem:assortlowerbound}, we know that
\[
\ALG\geq \sum_{\bs\in\mathcal{S}}p_1(\bs)\cdot\sum_{j\in[n]}\nu^1_j(\bs)\]
with $\nu^t_j(\bs)$ defined in \eqref{eqn:052202} for for any $j\in[n]$, $t\in[T]$, and any $\bs\in\mathcal{S}$. Moreover, from \Cref{lem:AssortUB} and \Cref{lem:assortupperbound}, we know that
\[
\OPT\leq \text{LP~}\eqref{lp:AssortADP}\leq(1+L)\cdot \sum_{\bs\in\mathcal{S}} p_1(\bs)\cdot\sum_{j\in[n]}\nu^1_j(\bs).
\]
Our proof is thus completed.

\end{myproof}

\begin{myproof}[Proof of \Cref{lem:assortlowerbound}]
We decompose the total reward collected by our \Cref{alg:assortBMP}, denoted by $\pi$, into two parts, and we have
\begin{equation}\label{eqn:052906}
\begin{aligned}
\sum_{t\in[T]}\sum_{j\in A_t(\bs_t)} \bI_{\{j\sim A_t(\bs_t)\}}\cdot r_{j}=&\underbrace{\sum_{t\in[T]} \sum_{j\in A_t(\bs_t)} \bI_{\{j\sim A_t(\bs_t)\}}\cdot \left[ r_{j}-\sum_{\bs'\in\mathcal{S}}p_t(\bs_t, \bs')\cdot\sum_{i\in\mathcal{A}_{j}}\frac{1}{C_i}\cdot\sum_{j'\in\mathcal{B}_i}\nu^{t+1}_{j'}(\bs') \right]}_{\text{I}}\\
&+\underbrace{\sum_{t\in[T]} \sum_{j\in A_t(\bs_t)} \bI_{\{j\sim A_t(\bs_t)\}}\cdot \left( \sum_{\bs'\in\mathcal{S}}p_t(\bs_t, \bs')\cdot\sum_{i\in\mathcal{A}_{j}}\frac{1}{C_i}\cdot\sum_{j'\in\mathcal{B}_i}\nu^{t+1}_{j'}(\bs') \right)}_{\text{II}},
\end{aligned}
\end{equation}
where $\bI_{\{j\sim A_t(\bs_t)\}}$ denotes whether or not the customer $t$ choose the product $j$ to purchase given the assortment $A_t(\bs_t)$.
We now analyze term I and term II separately.

\noindent \textbf{Bound term I.} From the decision rule in \Cref{alg:assortBMP} and more specifically, from the definition of $A_t(\bs_t)$ in \eqref{eqn:050702}, we have
\begin{align}
\text{term I}&=\sum_{t\in[T]}\sum_{j\in A_t(\bs_t)} \bI_{\{j\sim A_t(\bs_t)\}}\cdot \left[ r_{j}-\sum_{\bs'\in\mathcal{S}}p_t(\bs_t, \bs')\cdot\sum_{i\in\mathcal{A}_{j}}\frac{1}{C_i}\cdot\sum_{j'\in\mathcal{B}_i}\nu^{t+1}_{j'}(\bs') \right]\nonumber\\
&=\sum_{t\in[T]}\sum_{j\in A_t(\bs_t)} \bI_{\{j\sim A_t(\bs_t)\}}\cdot\bI_{\{\bc_t\geq \ba_{j}\}}\cdot \left[ r_{j}-\sum_{\bs'\in\mathcal{S}}p_t(\bs_t, \bs')\cdot\sum_{i\in\mathcal{A}_{j}}\frac{1}{C_i}\cdot\sum_{j'\in\mathcal{B}_i}\nu^{t+1}_{j'}(\bs') \right]^+.\label{eqn:052907}
\end{align}
Then, noting that the randomization $j\sim A_t(\bs_t)$ is independent of the randomization of $\bs_t$ and $\bc_t$, and also from the substitutability condition in \Cref{assump:assort}, we know that
\begin{equation}\label{eqn:050703}
\begin{aligned}
    \mathbb{E}[\text{term I}]
    &\geq \sum_{t\in[T]}\sum_{\bs\in\mathcal{S}}P(\bs_t=\bs)\cdot\sum_{j\in \hat{A}_t(\bs)}\phi_j(\hat{A}_t(\bs), \bs)\cdot \left(\left[ r_{j}-\sum_{\bs'\in\mathcal{S}}p_t(\bs, \bs')\cdot\sum_{i\in\mathcal{A}_{j}}\frac{1}{C_i}\cdot\sum_{j'\in\mathcal{B}_i}\nu^{t+1}_{j'}(\bs') \right]^+\cdot P(\bc_t\geq \ba_{j})   \right)
\end{aligned}
\end{equation}

\noindent \textbf{Bound term II.} From the bid price computing rule \eqref{eqn:052202}, for any $t\in[T]$, $j\in[n]$, $\bs\in\mathcal{S}$, we know that
\[
\nu^t_j(\bs)=\sum_{t'=t}^{T}\sum_{\bs'\in\mathcal{S}} P( \bs_{t'}=\bs'|\bs_t=\bs)\cdot\sum_{j\in\hat{A}_t(\bs')}\phi_j(\hat{A}_t(\bs'), \bs')\cdot\left[r_j-\sum_{\bs''\in\mathcal{S}}p_{t'}(\bs', \bs'')\cdot\sum_{i\in\mathcal{A}_j}\frac{1}{C_i}\cdot\sum_{j'\in\mathcal{B}_i}\nu^{t'+1}_{j'}(\bs'')\right]^+,
\]
where we use $P(\bs_{t'}=\bs'|\bs_t=\bs)$ to denote the probability that conditional on $\bs_t=\bs$, $\bs_{t'}$ is realized as $\bs'$. For each period $t\in[T]$, we also denote by $\bm{b}_t=\bm{C}-\bc_t$ the units of resources consumed at the beginning of period $t$. Then, by switching the sums and from the substitutability condition in \Cref{assump:assort}, we have
\[\begin{aligned}
&\mathbb{E}[\text{term II}]\\
=&\mathbb{E}\left[ \sum_{t\in[T]} \sum_{j\in A_t(\bs_t)} \bI_{\{j\sim A_t(\bs_t)\}}\cdot\bI_{\{\bc_t\geq \ba_{j}\}}\cdot \left( \sum_{\bs'\in\mathcal{S}}p_t(\bs_t, \bs')\cdot\sum_{i\in\mathcal{A}_{j}}\frac{1}{C_i}\cdot\sum_{j'\in\mathcal{B}_i}\nu^{t+1}_{j'}(\bs') \right) \right]\\
\geq&\mathbb{E}\left[ \sum_{t\in[T]} \sum_{j\in \hat{A}_t(\bs_t)} \bI_{\{j\sim \hat{A}_t(\bs_t)\}}\cdot\bI_{\{\bc_t\geq \ba_{j}\}}\cdot \left( \sum_{\bs'\in\mathcal{S}}p_t(\bs_t, \bs')\cdot\sum_{i\in\mathcal{A}_{j}}\frac{1}{C_i}\cdot\sum_{j'\in\mathcal{B}_i}\nu^{t+1}_{j'}(\bs') \right) \right]\\
=&\mathbb{E}\left[ \sum_{t\in[T]}\sum_{j\in \hat{A}_t(\bs_t)}\phi_j(\hat{A}_t(\bs_t), \bs_t)\left(\left[ r_{j}-\sum_{\bs'\in\mathcal{S}}p_t(\bs_t, \bs')\sum_{i\in\mathcal{A}_{j}}\frac{1}{C_i}\sum_{j'\in\mathcal{B}_i}\nu^{t+1}_{j'}(\bs') \right]^+ \sum_{i\in\mathcal{A}_{j}}\frac{1}{C_i}\sum_{t'=1}^{t-1}\bI_{\{j'\text{~purchased at time~}t', i\in\mathcal{A}_{j'}\}} \right)\right]\\
=& \sum_{t\in[T]}\sum_{\bs\in\mathcal{S}}P(\bs_t=\bs)\cdot\sum_{j\in \hat{A}_t(\bs)}\phi_j(\hat{A}_t(\bs), \bs)\cdot\left(\left[ r_{j}-\sum_{\bs'\in\mathcal{S}}p_t(\bs, \bs')\cdot\sum_{i\in\mathcal{A}_{j}}\frac{1}{C_i}\cdot\sum_{j'\in\mathcal{B}_i}\nu^{t+1}_{j'}(\bs') \right]^+\cdot \sum_{i\in\mathcal{A}_{j}}\frac{\mathbb{E}[b_{i,t}|\bs_t=\bs]}{C_i}\right).
\end{aligned}\]
We apply Markov's inequality to show that for each $i\in[m]$, it holds
\[
\frac{\mathbb{E}[b_{i,t}|\bs_t=\bs]}{C_i}\geq P(b_{i,t}\geq C_i|\bs_t=\bs)=P(c_{i,t}\leq 0|\bs_t=\bs).
\]
Then, by the union bound, we have
\[
\sum_{i\in\mathcal{A}_{j}}\frac{\mathbb{E}[b_{i,t}|\bs_t=\bs]}{C_i}\geq \sum_{i\in\mathcal{A}_{j}}P(c_{i,t}\leq 0|\bs_t=\bs) \geq P\left(\exists i\in\mathcal{A}_{j}: c_{i,t}< a_{i,j}|\bs_t=\bs\right)=1-P(\bc_t\geq \ba_{j}|\bs_t=\bs).
\]
Therefore, we know that
\begin{equation}\label{eqn:053001}
\begin{aligned}
&\mathbb{E}[\text{term II}]\\
=&\sum_{t\in[T]}\sum_{\bs\in\mathcal{S}}P(\bs_t=\bs)\cdot\sum_{j\in \hat{A}_t(\bs)}\phi_j(\hat{A}_t(\bs), \bs)\cdot\left(\left[ r_{j}-\sum_{\bs'\in\mathcal{S}}p_t(\bs, \bs')\cdot\sum_{i\in\mathcal{A}_{j}}\frac{1}{C_i}\cdot\sum_{j'\in\mathcal{B}_i}\nu^{t+1}_{j'}(\bs') \right]^+\cdot \sum_{i\in\mathcal{A}_{j}}\frac{\mathbb{E}[b_{i,t}|\bs_t=\bs]}{C_i}\right)\\
\geq& \sum_{t\in[T]}\sum_{\bs\in\mathcal{S}}P(\bs_t=\bs)\cdot\sum_{j\in \hat{A}_t(\bs)}\phi_j(\hat{A}_t(\bs), \bs)\cdot\left(\left[ r_{j}-\sum_{\bs'\in\mathcal{S}}p_t(\bs, \bs')\cdot\sum_{i\in\mathcal{A}_{j}}\frac{1}{C_i}\cdot\sum_{j'\in\mathcal{B}_i}\nu^{t+1}_{j'}(\bs') \right]^+\cdot  \left( 1-P(\bc_t\geq \ba_{j}|\bs_t=\bs) \right)   \right).
\end{aligned}
\end{equation}
Combining \eqref{eqn:052906}, \eqref{eqn:050703}, and \eqref{eqn:053001}, we have
\begin{equation}\label{eqn:053003}
\begin{aligned}
&\mathbb{E}\left[\sum_{t\in[T]}\sum_{j\in A_t(\bs_t)} \bI_{\{j\sim A_t(\bs_t)\}}\cdot r_{j}\right]\\
\geq&\sum_{t\in[T]}\sum_{\bs\in\mathcal{S}}P(\bs_t=\bs)\cdot\sum_{j\in \hat{A}_t(\bs)}\phi_j(\hat{A}_t(\bs), \bs)\cdot\left(\left[ r_{j}-\sum_{\bs'\in\mathcal{S}}p_t(\bs, \bs')\cdot\sum_{i\in\mathcal{A}_{j}}\frac{1}{C_i}\cdot\sum_{j'\in\mathcal{B}_i}\nu^{t+1}_{j'}(\bs') \right]^+  \right).
\end{aligned}
\end{equation}
Finally, from the bid price computing rule \eqref{eqn:052202} and by noting $\nu^{T+1}_j(\bs)=0$ for any $j\in[n]$ and $\bs\in\mathcal{S}$, we also have
\begin{equation}\label{eqn:053004}
 \sum_{\bs\in\mathcal{S}}p_1(\bs)\cdot \sum_{j\in[n]}\nu^1_j(\bs)=  \sum_{t\in[T]}\sum_{\bs\in\mathcal{S}}P(\bs_t=\bs)\cdot\sum_{j\in \hat{A}_t(\bs)}\phi_j(\hat{A}_t(\bs), \bs)\cdot\left(\left[ r_{j}-\sum_{\bs'\in\mathcal{S}}p_t(\bs, \bs')\cdot\sum_{i\in\mathcal{A}_{j}}\frac{1}{C_i}\cdot\sum_{j'\in\mathcal{B}_i}\nu^{t+1}_{j'}(\bs') \right]^+  \right).
\end{equation}
Comparing \eqref{eqn:053003} and \eqref{eqn:053004}, we have
\[
\mathbb{E}\left[\sum_{t\in[T]}\sum_{j\in A_t(\bs_t)} \bI_{\{j\sim A_t(\bs_t)\}}\cdot r_{j}\right]\geq \sum_{\bs\in\mathcal{S}}p_1(\bs)\cdot \sum_{j\in[n]}\nu^1_j(\bs)
\]
which completes our proof.

\end{myproof}

\begin{myproof}[Proof of \Cref{lem:assortupperbound}]
We first show that $\{\hat{\beta}^t_i(\bs), \hat{\theta}^t(\bs), \forall i\in[m], \forall t\in[T], \forall \bs\in\mathcal{S}\}$ defined in \eqref{eqn:042301} and \eqref{eqn:042302} is feasible to LP \eqref{lp:AssortADP} with the bid price $\nu^t_j(\bs)$ defined in \eqref{eqn:052202}. Note that from the definition in \eqref{eqn:052202}, we must have $\nu^t_j(\bs)\geq\sum_{\bs'\in\mathcal{S}}p_t(\bs, \bs')\cdot\nu^{t+1}_j(\bs')$ for any $j\in[n]$, $t\in[T]$ and $\bs\in\mathcal{S}$, which implies that
\[
\hat{\beta}^t_i(\bs)\geq\sum_{\bs'\in\mathcal{S}}p_t(\bs, \bs')\cdot\hat{\beta}^{t+1}_i(\bs').
\]
Therefore, in order for the constraint \eqref{const:AssortADP} to be satisfied, it is sufficient to show that
\[
 \hat{\theta}^t(\bs)-\sum_{\bs'\in\mathcal{S}}p_t(\bs, \bs')\cdot\hat{\theta}^{t+1}(\bs')\geq \sum_{j\in A}\phi_j(A, \bs)\cdot \left[ r_{j}-\sum_{\bs'\in\mathcal{S}}p_t(\bs, \bs')\cdot\sum_{i\in[m]}a_{i,j}\cdot \hat{\beta}^{t+1}_{i}(\bs') \right]^+,
\]
holds for any $\forall A\in\mathcal{F}$.
Note that
\[\begin{aligned}
\hat{\theta}^t(\bs)-\sum_{\bs'\in\mathcal{S}}p_t(\bs, \bs')\cdot\hat{\theta}^{t+1}(\bs')&=\sum_{j\in[n]}\nu^t_j(\bs)-\sum_{\bs'\in\mathcal{S}}p_t(\bs, \bs')\cdot\sum_{j\in[n]}\nu^t_j(\bs')\\
&=\sum_{j\in\hat{A}_t(\bs)}\phi_j(\hat{A}_t(\bs), \bs)\cdot \left[r_{j}-\sum_{\bs'\in\mathcal{S}}p_t(\bs, \bs')\cdot\sum_{i\in\mathcal{A}_{j}}\frac{1}{C_i}\cdot\sum_{j'\in\mathcal{B}_i}\nu^{t+1}_{j'}(\bs') \right]^+\\
&\geq \sum_{j\in A}\phi_j(A, \bs)\cdot \left[r_{j}-\sum_{\bs'\in\mathcal{S}}p_t(\bs, \bs')\cdot\sum_{i\in\mathcal{A}_{j}}\frac{1}{C_i}\cdot\sum_{j'\in\mathcal{B}_i}\nu^{t+1}_{j'}(\bs') \right]^+\\
&=\sum_{j\in A}\phi_j(A, \bs)\cdot
\left[r_{j}-\sum_{\bs'\in\mathcal{S}}p_t(\bs, \bs')\cdot\sum_{i\in\mathcal{A}_{j}}\beta^{t+1}_{i}(\bs') \right]^+\\
&=\sum_{j\in A}\phi_j(A, \bs)\cdot\left[ r_{j}-\sum_{\bs'\in\mathcal{S}}p_t(\bs, \bs')\cdot\sum_{i\in[m]}a_{i,j}\cdot \hat{\beta}^{t+1}_{i}(\bs') \right]^+
\end{aligned}\]
for all $A\in\mathcal{F}$, where the first inequality follows from the definition of $\hat{A}_t(\bs)$ in \eqref{eqn:050701}.
Therefore, we show that $\{\hat{\beta}^t_i(\bs), \hat{\theta}^t(\bs), \forall i\in[m], \forall t\in[T], \forall \bs\in\mathcal{S}\}$ defined in \eqref{eqn:042301} and \eqref{eqn:042302} with the bid price $\nu^t_j(\bs)$ defined in \eqref{eqn:052202} is feasible to LP \eqref{lp:AssortADP}.

The objective value of LP \eqref{lp:AssortADP} under the feasible solution $\{\hat{\beta}^t_i(\bs), \hat{\theta}^t(\bs), \forall i\in[m], \forall t\in[T], \forall \bs\in\mathcal{S}\}$ can be upper bounded as follows.
\[\begin{aligned}
\sum_{\bs\in\mathcal{S}} p_1(\bs)\cdot \left( \hat{\theta}^1(\bs)+\sum_{i\in[m]}C_i\cdot\hat{\beta}^1_{i}(\bs) \right)&=\sum_{\bs\in\mathcal{S}} p_1(\bs)\cdot \sum_{j\in[n]}\nu^1_j(\bs)+\sum_{\bs\in\mathcal{S}}p_1(\bs)\cdot \sum_{i\in[m]}\sum_{j\in\mathcal{B}_i} \nu^t_j(\bs) \\
&=\sum_{\bs\in\mathcal{S}} p_1(\bs)\cdot \sum_{j\in[n]}\nu^1_j(\bs)+\sum_{\bs\in\mathcal{S}}p_1(\bs)\cdot \sum_{j\in[n]}\sum_{i\in\mathcal{A}_j} \nu^t_j(\bs)\\
&\leq (1+L)\cdot \sum_{\bs\in\mathcal{S}} p_1(\bs)\cdot \sum_{j\in[n]}\nu^1_j(\bs)
\end{aligned}\]
where it follows from definition \eqref{eqn:042203} that the number of elements in the set $\mathcal{A}_j$ is upper bounded by $L$ for any $j\in[n]$. Our proof is thus completed.

\end{myproof}

\end{APPENDICES}

\end{document}